\documentclass[aps,prx,reprint,amsmath,amssymb,floatfix,superscriptaddress]{revtex4-2}

\usepackage{graphicx}
\usepackage{placeins}

\newcommand{\Pfinal}{P_{\mathrm{final}}}
\newcommand{\Pbar}{\bar{P}_{A}}
\newcommand{\Pbardeph}{\bar{P}_{A}^{\mathrm{deph}}}
\newcommand{\Pexc}{P_{\mathrm{excess}}^{\phi}}
\newcommand{\Ptwo}{\Pfinal^{\ket{2}}}
\newcommand{\Pthree}{\Pfinal^{\ket{3}}}
\newcommand{\Tphi}{T_{\phi}}
\newcommand{\dQ}{d_{Q}}
\newcommand{\ket}[1]{|#1\rangle}
\newcommand{\Rx}{R_{X}(\pi/2)}

\begin{document}

\title{Separate Control of Transient Leakage Exposure and Endpoint Leakage \\in Fast Transmon Gates}

\author{Haoran Yang}
\email{19561362267@163.com}
\author{Fudong Liu}
\email{liufudong@email.edu}
\author{Weilong Wang}
\author{Yangyang Fei}
\author{Zheng Shan}
\email{shanzheng@email.edu}
\affiliation{Information Engineering University, Zhengzhou, Henan, China}

\begin{abstract}
Leakage to noncomputational states limits the speed of single-qubit gates in
weakly anharmonic transmons. Conventional pulse-shaping methods, including
derivative removal by adiabatic gate (DRAG), primarily suppress the leakage
remaining at the end of a gate. Here we demonstrate that endpoint leakage 
and the transient leakage population that accumulates during the gate 
represent distinct control objectives. Endpoint leakage is associated with the drive spectrum at the
anharmonicity, whereas transient exposure depends on spectral weight over a
finite frequency band and governs the additional leakage induced by dephasing.
A spectral null at the leakage transition therefore suppresses the endpoint
amplitude without necessarily reducing transient exposure. Based on this
distinction, we introduce a path--endpoint separation pulse that combines
transient-path shaping with a two-tone endpoint correction. For a $10$~ns
$R_X(\pi/2)$ gate with an anharmonicity magnitude of $0.2$~GHz, numerical simulations show
a $21\%$ reduction in transient exposure relative to cosine DRAG and a
corresponding $20\%$ reduction in dephasing-induced excess leakage. The two
correction tones further suppress residual leakage through the $\ket{2}$ and
$\ket{3}$ channels, lowering the coherent endpoint leakage from approximately
$7\times10^{-7}$ to $3\times10^{-8}$ without increasing transient exposure.
These results establish transient exposure and endpoint leakage as
complementary targets for the design of fast transmon gates.
\end{abstract}

\maketitle

% ============================================================================
\section{Introduction}\label{sec:intro}
% ============================================================================

Transmon qubits are weakly anharmonic oscillators whose
$\ket{0}\!\leftrightarrow\!\ket{1}$ and $\ket{1}\!\leftrightarrow\!\ket{2}$
transitions are separated only by the anharmonicity $\eta$ (typically
${\sim}200$~MHz)~\cite{Koch2007,Blais2021,Krantz2019,Kjaergaard2020}. 
Shortening the gate time requires stronger, broader-band
microwave drives whose spectra overlap more strongly with the leakage transition, so
fast gates must control population leaving the computational subspace in
addition to optimizing the in-subspace fidelity~\cite{Motzoi2009,Chen2016}.
Because leakage transfers population outside the computational subspace, it is not
captured by the Pauli error channels of standard error-correction analyses and is
especially detrimental to fault-tolerant operation~\cite{Wood2018,Fowler2012,McEwen2021,Miao2023}.
Leakage, however, is not fully characterized by the population remaining
outside the computational subspace at the end of a gate. Even when the
closed-system endpoint leakage vanishes, a fast drive can transiently populate
or coherently admix noncomputational states, producing a finite
time-integrated leakage exposure. This distinction matters because, as shown
by Strauch~\cite{Strauch2025}, pure dephasing converts such transient exposure
into residual leakage even if the closed-system evolution returns entirely to
the computational subspace at the end of the gate.

Methods for suppressing leakage span a hierarchy of waveform
parametrizations. Physics-motivated parametric approaches reduce the control complexity 
by restricting the waveform to a compact ansatz with a small number of tunable parameters. 
These parameters may be obtained analytically, selected heuristically, or calibrated experimentally.
Derivative-based and Fourier-shaped envelopes, including derivative removal
by adiabatic gate (DRAG), Fourier ansatz spectrum tuning (FAST) DRAG, and
higher-derivative (HD) DRAG, suppress spectral weight at or around the leakage
transition~\cite{Motzoi2009,Gambetta2011,Lucero2010,Hyyppa2024}. Active
leakage cancellation (ALC) instead augments the primary drive with a calibrated
auxiliary tone near the leakage transition, canceling the accumulated
leakage amplitude through destructive interference and achieving hardware
leakage below $10^{-5}$~\cite{Chiaro2025}. Numerical optimal-control methods
determine some or all waveform parameters by directly minimizing a
gate-performance objective~\cite{Khaneja2005,Werninghaus2021}. At one extreme,
sample-wise parametrizations optimize a large number of independent control
variables and provide high flexibility within the chosen time discretization.

Between compact analytic constructions and
sample-wise control, parametrizations based on truncated smooth bases, such as
a chopped Fourier basis~\cite{Caneva2011}, reduce the number of independent
variables while retaining considerable flexibility. Such reduced-basis
parametrizations have achieved strong hardware performance: a seven-parameter
Fourier-series pulse recently achieved $99.9\%$ controlled-$Z$ fidelity and
outperformed a sample-wise parametrization of the same pulse
family~\cite{Glaser2025}.
Model-based open-loop solutions may be sensitive to inaccuracies in the
Hamiltonian model and control-transfer function. Experimental closed-loop
optimization can compensate for such model mismatch, but optimization of the
full parameter set may incur substantial calibration cost. It can also make
performance gains difficult to attribute to specific spectral or dynamical
mechanisms~\cite{Werninghaus2021}.
Despite this diversity in waveform construction, many leakage-suppression
schemes are designed or evaluated primarily in terms of the leakage remaining
at gate completion. In the perturbative spectral picture, DRAG-type and
Fourier-based shaping suppress spectral weight near the leakage transition 
to reduce the final transition amplitude; ALC imposes an
interference condition that cancels the accumulated endpoint amplitude; and
endpoint-focused optimal-control implementations penalize the final leakage
or gate infidelity. These design criteria may influence the transient
dynamics indirectly, but they do not explicitly minimize the
time-integrated leakage exposure accumulated during the gate. This exposure
is a physically distinct quantity and, as we show, determines the leading
dephasing-induced leakage. Increasing the flexibility of the waveform
parametrization does not by itself guarantee its suppression: when the
objective emphasizes only the final leakage, optimized solutions can reach a
very low endpoint value while retaining substantial noncomputational exposure
during the gate, a behavior we demonstrate explicitly with an endpoint-only optimum
in Sec.~\ref{sec:coherent}. Reinforcement-learning gate design exhibits the same
behavior, with high-fidelity solutions exploiting large transient excursions
that return to low leakage at gate completion~\cite{Nguyen2024}.

Such excursions become a liability under dephasing; 
we therefore treat transient exposure and residual endpoint leakage 
as separate control objectives. We implement this separation
through a staged protocol termed \emph{path--endpoint separation}, whose
resulting composite waveform is the \emph{path--endpoint separation pulse}
(PESP). A compact \emph{path-shaping pulse} (PSP) first minimizes the
time-integrated leakage exposure, after which an
\emph{endpoint-cancellation pulse} (ECP) suppresses the residual endpoint
leakage amplitude. A final amplitude recalibration, together with a virtual-$Z$
correction, completes the design. The PSP controls the extent to which
population transiently occupies the leakage states during the gate, whereas
the ECP suppresses the residual leakage amplitude remaining at gate completion.
By assigning these two physical tasks to separate pulse components and
optimization stages, PESP suppresses both transient exposure and endpoint
leakage while allowing the contribution of each stage to be isolated directly
through ablation, rather than inferred from a joint optimum over many
parameters.

In the frequency-domain picture, endpoint leakage is
associated with the drive spectrum at the anharmonicity, whereas transient
exposure depends on spectral weight over a finite band around it. Guided by this distinction, 
PESP combines path shaping with a two-tone endpoint correction. Four-level
simulations reveal an operating-point trade-off: 
increasing the weight assigned to the path-exposure penalty
initially suppresses the $\ket{2}$ back-action but eventually enhances the
$\ket{2}\!\to\!\ket{3}$ cascade. At the resulting path-exposure knee, the
two-tone endpoint correction lowers the coherent leakage from approximately
$7\times10^{-7}$ to $3\times10^{-8}$ without appreciably changing the
transient exposure. Independently, path shaping reduces the dephasing exposure
by approximately $21\%$ relative to cosine DRAG and reduces the excess leakage in
Lindblad simulations by approximately $20\%$. These results establish the
spectral distinction between transient exposure and endpoint leakage and
provide a mechanism-resolved strategy for controlling them separately.

The present study is numerical and is intended to establish the distinction
between transient exposure and endpoint leakage rather than to claim a
hardware advantage over fully calibrated alternatives.

% ============================================================================
\section{Model and metrics}\label{sec:model}
% ============================================================================

\subsection{Hamiltonian and simulation model}

We use a rotating-frame Kerr-oscillator truncation,
\begin{equation}
H(t)=-\frac{\eta}{2}\,n(n-1)
+\frac{1}{2}\!\left[\Omega(t)\,a+\Omega^{*}(t)\,a^{\dagger}\right],
\label{eq:H}
\end{equation}
where $\eta>0$ is the magnitude of the negative transmon anharmonicity,
$\eta/2\pi\approx 200$~MHz, $a$ is the annihilation operator, and
$n=a^\dagger a$. Three-level calculations include the leakage level $\ket{2}$ and are used for
pulse optimization and mechanism diagnostics, including the endpoint-only
optimum of Sec.~\ref{sec:coherent}. Four-level calculations include $\ket{2}$
and $\ket{3}$ and determine the endpoint floor, its dependence on
$w_{\mathrm{path}}$, and the operating point. Results obtained with the two
truncations are identified explicitly. Unless a Lindblad calculation is explicitly
specified, all simulations are closed-system unitary evolutions.

\subsection{Target gate and projected fidelity}

The target is $\Rx$. 
We report the computational-subspace projected average gate
fidelity, maximized analytically over a post-gate virtual-$Z$~\cite{McKay2017},
\begin{equation}
F_{\mathrm{comp}}=\max_{Z}\,
\frac{\operatorname{Tr}(M^\dagger M)
+\big|\operatorname{Tr}(U_{\mathrm{target}}^\dagger Z M)\big|^2}{d(d+1)},
\label{eq:Fcomp}
\end{equation}
where $M=P_Q U(T) P_Q$ and $d=2$. The maximization gives
$\max_Z|\operatorname{Tr}(\cdot)|=|A|+|B|$, with
$A=U_{00}^{*}M_{00}+U_{01}^{*}M_{01}$ and
$B=U_{10}^{*}M_{10}+U_{11}^{*}M_{11}$. Thus, the reported infidelity excludes phase errors that can be corrected by
a post-gate virtual-$Z$ operation.

\subsection{Leakage metrics}

We define the endpoint leakage and the time-averaged path exposure as
\begin{align}
\Pfinal &=\frac{1}{\dQ}\sum_{j\in Q,\,k\in A}
\big|\langle k|U(T)|j\rangle\big|^2 ,\label{eq:pfinal}\\
\Pbar &=\frac{1}{T \dQ}\sum_{j\in Q,\,k\in A}
\int_0^T \big|\langle k|U(t)|j\rangle\big|^2\,dt .
\label{eq:path}
\end{align}
We write $P_A(t)=\dQ^{-1}\sum_{j\in Q,\,k\in A}|\langle k|U(t)|j\rangle|^2$ for
the instantaneous leakage population, so that $\Pbar$ is its time average; its
maximum over the gate enters the peak-leakage penalty of Eq.~(\ref{eq:cmain}).
A pulse can have a very small $\Pfinal$ yet a large $\Pbar$. The
Strauch-inspired dephasing coupling and its path average are
\begin{align}
A_{j\to k}(t)&=\langle a_k|U^\dagger(t)\,n\,U(t)|q_j\rangle ,\nonumber\\
\Pbardeph&=\frac{1}{T\dQ}\sum_{j,k}\int_0^T|A_{j\to k}(t)|^2\,dt .
\label{eq:deph}
\end{align}
The dephasing-induced excess leakage is
$\Pexc=P_{\mathrm{leak,noisy}}(T)-P_{\mathrm{leak,ideal}}(T)$, which isolates the
noise contribution from the closed-system floor.

\subsection{First-order dephasing relation}

Pure dephasing is modeled by the
Lindblad equation
\begin{equation}
\dot\rho=-i[H,\rho]
+\gamma_\phi\!\left(n\rho n-\tfrac{1}{2}\{n^{2},\rho\}\right),
\label{eq:lindblad}
\end{equation}
under which the $\ket{0}$--$\ket{1}$ coherence decays as
$e^{-\gamma_\phi t/2}$, so that $\gamma_\phi=2/\Tphi$. First-order
perturbation theory gives
\begin{equation}
\Pexc \approx \gamma_\phi \int_0^T \frac{1}{\dQ}\sum_{j,k}|A_{j\to k}(t)|^2\,dt
= \gamma_\phi\,T\,\Pbardeph .
\label{eq:firstorder}
\end{equation}
We use $\Pbar$ [Eq.~(\ref{eq:path})] as the Stage-1 design proxy because it
agrees with $\Pbardeph$ to within ${\lesssim}2\%$ in the main Pareto region
(Appendix~\ref{app:proxy}). Its physical relevance is assessed independently
through the four-level Lindblad excess leakage $\Pexc$, which is an output of
the noisy dynamics and is not included in any optimization objective;
Eq.~(\ref{eq:firstorder}) is verified quantitatively in
Sec.~\ref{sec:dephasing}, where the endpoint-only baseline serves as a control
constructed to minimize $\Pfinal$ without penalizing $\Pbardeph$.

\subsection{Spectral and hardware metric}

We define the high-frequency energy fraction $C_{\mathrm{hf}}$ as the fraction of
rotating-frame envelope spectral power above
$|\omega|>2\pi\times0.8$~GHz. This cutoff represents the assumed
digital-to-analog converter and mixer bandwidth.

% ============================================================================
\section{Frequency-domain mechanism: endpoint versus path}\label{sec:mechanism}
% ============================================================================

A first-order perturbative analysis casts the endpoint--path distinction in the
frequency domain, where the two metrics become manifestly different functionals
of the same control spectrum. In the displaced/adiabatic frame, leakage out of
the computational subspace is dominated by the
$\ket{1}\!\leftrightarrow\!\ket{2}$ channel. Treating the effective coupling
$\Lambda(t)$ (the envelope together with its DRAG correction) to first order and
rotating at the leakage detuning $\eta$, the amplitude accumulated in $\ket{2}$
up to time $t$ is the running Fourier integral
\begin{equation}
c_2(t)\simeq -i\int_0^t \Lambda(t')\,e^{i\eta t'}\,dt' .
\label{eq:c2}
\end{equation}
Both leakage metrics are quadratic in $c_2$, but they sample it differently. The
endpoint leakage evaluates it once, at $t=T$,
\begin{equation}
\Pfinal \propto |c_2(T)|^2 = \big|\tilde\Lambda(\eta)\big|^2 ,\qquad
\tilde\Lambda(\eta)\equiv\int_0^T\!\Lambda(t)\,e^{i\eta t}\,dt ,
\label{eq:final_spectral}
\end{equation}
i.e., the squared drive-spectrum weight at the leakage detuning $\eta$, a single
spectral sample. The path leakage is instead the time average of the running
population,
\begin{equation}
\Pbar \propto \frac{1}{T}\int_0^T |c_2(t)|^2\,dt .
\label{eq:path_spectral}
\end{equation}
To expose its spectral content, expand the modulus with Eq.~(\ref{eq:c2}),
\begin{equation}
|c_2(t)|^2=\int_0^t\!\!\int_0^t \Lambda(s)\Lambda^*(s')\,e^{i\eta(s-s')}\,ds\,ds' ,
\label{eq:c2sq}
\end{equation}
insert it into Eq.~(\ref{eq:path_spectral}), and exchange the order of
integration. For fixed $(s,s')$ the outer time obeys $t\ge\max(s,s')$, so the
$t$-integration gives $\int_{\max(s,s')}^{T}dt=T-\max(s,s')$ and
\begin{equation}
\Pbar \propto \frac{1}{T}\int_0^T\!\!\int_0^T\!\Lambda(t)\Lambda^*(t')\,
e^{i\eta(t-t')}\,\big[\,T-\max(t,t')\,\big]\,dt\,dt' .
\label{eq:path_kernel}
\end{equation}
The kernel admits the Gram representation
\begin{multline}
T-\max(t,t')=\min(T-t,\,T-t')\\
=\int_0^T\Theta(u-t)\,\Theta(u-t')\,du ,
\label{eq:kernel}
\end{multline}
with $\Theta$ the step function, so Eq.~(\ref{eq:path_kernel}) is a manifestly
positive-semidefinite quadratic form in $\Lambda$, as it must be. Equations
(\ref{eq:final_spectral}) and (\ref{eq:path_kernel}) are the two functionals: a
point evaluation of the control spectrum versus a kernel-weighted integral of it.

The frequency-domain meaning follows on writing $h(t)\equiv\Lambda(t)\,e^{i\eta
t}$, so that $c_2(t)=-i\int_0^t h\,dt'$ and the endpoint
$\tilde\Lambda(\eta)=\hat h(0)$ is the zero-frequency (net-area) component of
$h$; the endpoint fixes this component alone. The time average of a running
integral, by contrast, weights the low-frequency content of $h$ over a band of
width ${\sim}1/T$. Splitting the kernel of Eq.~(\ref{eq:path_kernel}) into its
stationary and non-stationary parts (Appendix~\ref{app:fejer}) gives the exact
representation
\begin{multline}
\frac{1}{T}\!\int_0^T\!|c_2(t)|^2\,dt
=\frac{1}{2}\!\int\!\frac{d\omega}{2\pi}\,\big|\tilde\Lambda(\omega)\big|^2\,
\mathcal{F}_T(\omega-\eta)+\mathcal{R},\\
\mathcal{F}_T(\nu)=\frac{1}{T}\!\left[\frac{\sin(\nu T/2)}{\nu/2}\right]^2 ,
\label{eq:fejer}
\end{multline}
with $\mathcal{F}_T$ a nonnegative Fej\'er weight of width ${\sim}2\pi/T$
centered on $\eta$ that integrates to unity,
$\int\!(d\omega/2\pi)\,\mathcal{F}_T=1$, and a remainder $\mathcal{R}$ that is
linear in the endpoint amplitude and vanishes identically when
$\tilde\Lambda(\eta)=0$ [Eq.~(\ref{eq:fejer_remainder})]. The spectral
representation is used for interpretation only; every reported path value is
computed in the time domain from Eq.~(\ref{eq:path}). Thus the endpoint
constrains the single frequency $\omega=\eta$, whereas the path constrains an
entire neighborhood of it: forcing $\tilde\Lambda(\eta)=0$ (a spectral-null condition at the leakage
transition) sets the
integrand of Eq.~(\ref{eq:fejer}) to zero only at the band center and does not
control the integral that fixes $\Pbar$. This is the analytic statement of the
mechanism separation.

The two objectives can even drive the same control coordinate in opposite
directions. At the central operating point the leakage transition coincides with
the second harmonic of the cosine envelope basis introduced in
Sec.~\ref{sec:ansatz} [$g_n(t)=1-\cos(\omega_n t)$, $\omega_n=2\pi n/T$],
$\eta=\omega_2$. Writing the in-phase envelope as $\Lambda(t)\propto\sum_n a_n
g_n(t)$ (with $a_1\equiv1$) and using the integer-period orthogonality relations
\begin{equation}
\int_0^T\! e^{i\omega_2 t}\,dt=0,\qquad
\int_0^T\!\cos(\omega_n t)\,e^{i\omega_2 t}\,dt=\tfrac{T}{2}\,\delta_{n,2} ,
\label{eq:ortho}
\end{equation}
only the $g_2$ term survives the endpoint sample,
\begin{equation}
\tilde\Lambda(\eta)=\tilde\Lambda(\omega_2)\propto -\tfrac{T}{2}\,a_2 ,
\label{eq:a2}
\end{equation}
so the in-phase endpoint weight is set by $a_2$ alone, with the quadrature
component controlled independently by $\beta_1$. Within this first-order
in-phase description and at $\eta=\omega_2$, the spectral-null condition
reduces to $a_2=0$, whereas minimizing the band integral of Eq.~(\ref{eq:fejer})
reshapes the running excursion of $c_2(t)$ and drives $a_2$ away from zero, at
the price of a nonzero $\tilde\Lambda(\eta)$ that the endpoint tone
subsequently suppresses. The path objective thus deliberately populates the
spectral component that a spectral-null construction removes. Imposing an
endpoint spectral null therefore does not by itself reduce the transient
exposure, consistent with the matched-budget FAST~DRAG baseline of
Sec.~\ref{sec:coherent}.

\subsection{Mechanism-defined control coordinates}

The staged parametrization
associates each optimized parameter with a specific part of the leakage
response before any numerical search is performed. At $\eta=\omega_2$ the
coefficient $a_2$ sets the in-phase spectral component at the leakage
transition within the first-order description above, and $\beta_1$ supplies
the leading DRAG quadrature. The auxiliary tones near $-\eta$ and $-2\eta$
act predominantly on the residual $\ket{2}$ and $\ket{3}$ endpoint
populations, respectively. These assignments are tested by ablation and by the
level-resolved results of Sec.~\ref{sec:coherent} (Table~\ref{tab:params}).
Sample-wise optimal control realizes the same physical functionals but does
not impose such a parameter-level assignment: a single sample amplitude
generally changes the waveform, its spectrum, and the leakage trajectory
simultaneously, so attribution requires a posteriori
analysis~\cite{Khaneja2005,Werninghaus2021}.

% ============================================================================
\section{Pulse ansatz}\label{sec:ansatz}
% ============================================================================

\subsection{Path-shaping pulse}

\begin{figure*}[t]
\centering
\includegraphics[width=\textwidth]{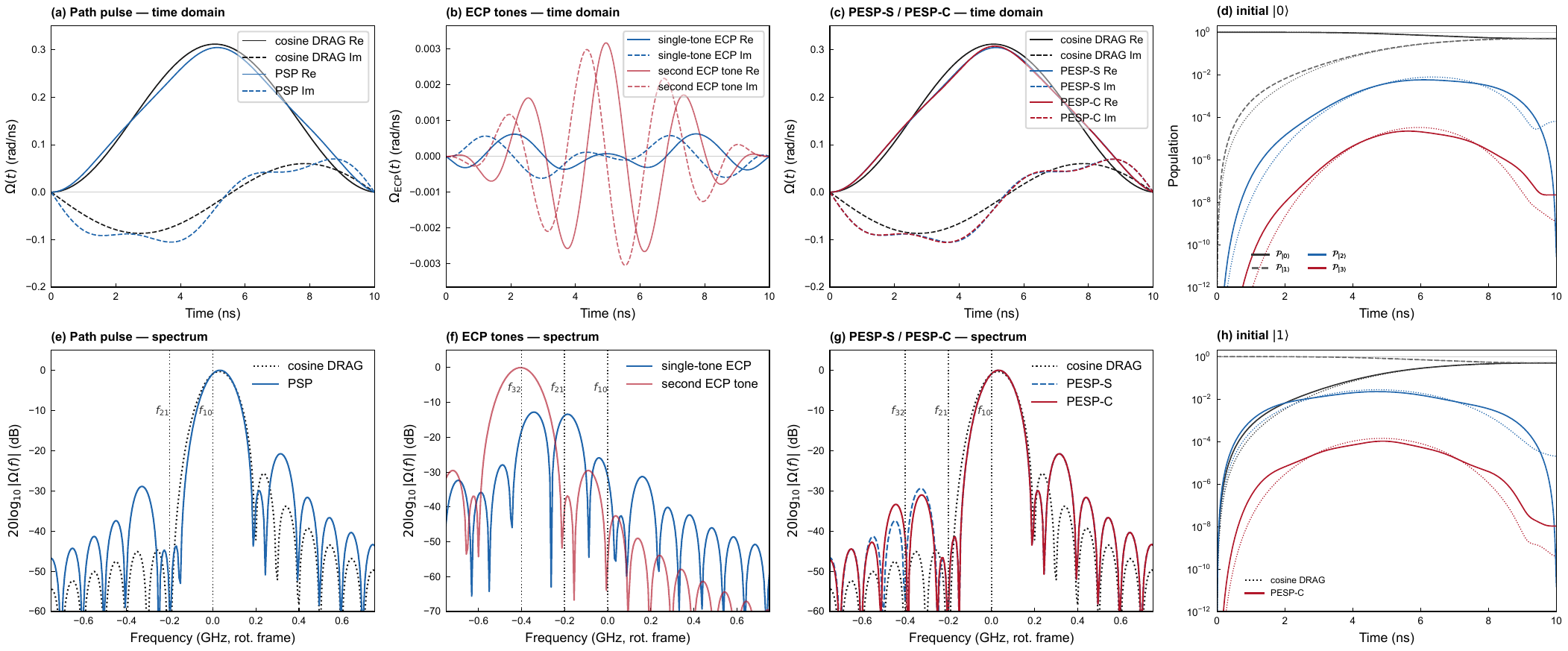}
\caption{\label{fig:waveform_combined}PESP construction: waveforms, spectra, and
leakage trajectories.
(a)~Time-domain drive waveform $\Omega(t)$ for cosine DRAG (gray dotted) and the
PSP (blue solid); real (solid) and imaginary (dashed) components.
(b)~Isolated endpoint-cancellation tones: single-tone ECP (blue) and second ECP
tone (red); real and imaginary components as in (a).
(c)~Time-domain waveforms for cosine DRAG (black dotted), PESP-S (blue dashed),
and PESP-C (red solid).
(d)~Leakage trajectories for initial $\ket{0}$: per-level populations
$\mathcal{P}_{\ket{0}}$, $\mathcal{P}_{\ket{1}}$, $\mathcal{P}_{\ket{2}}$,
$\mathcal{P}_{\ket{3}}$ for cosine DRAG (dotted) and PESP-C (solid).
(e)--(g)~Corresponding amplitude spectra $|\Omega(\omega)|$ (dB scale); dashed
vertical lines mark the computational frequency $f_{10}{=}0$ and the leakage
transitions $f_{21}{=}{-}\eta/2\pi$ and $f_{32}{=}{-}2\eta/2\pi$
[the latter only in (f),(g)].
(h)~Same as (d) for initial $\ket{1}$.}
\end{figure*}

The PSP envelope is built from cosine basis functions
$g_n(t)=1-\cos(2\pi n t/T)$,
\begin{equation}
F_\theta(t)=\mathcal{N}\big[g_1(t)+a_2 g_2(t)+a_3 g_3(t)\big],
\end{equation}
and the complex main drive applies a generalized-DRAG quadrature and an
AC-Stark carrier detuning $\delta_m$,
\begin{equation}
\Omega_{\mathrm{main}}(t)=A_m\left[F_\theta+i\beta_1\frac{\dot F_\theta}{\Delta}
+\beta_2\frac{\tilde F_2}{\Delta^2}\right]e^{i\delta_m t},
\label{eq:main}
\end{equation}
with $\Delta=-(\eta+\delta_m)$. The optimized parameters are
$\theta_{\mathrm{main}}=[a_2,a_3,\beta_1,\beta_2,\delta_m]$; the overall
amplitude $A_m$ is fixed by the gate-area constraint and a scalar amplitude
calibration and is not a free dimension. Cosine DRAG is recovered as the special case $a_2=a_3=0$, $\beta_1=1$,
$\beta_2=0$, with $\delta_m$ and the overall amplitude calibrated by the same
procedure as for the other pulse families.  The cosine parametrization is shared with the Fourier ansatz of FAST
DRAG~\cite{Hyyppa2024}; what distinguishes the PSP is the explicit time-domain
path-exposure penalty added in the staged cost (Sec.~\ref{sec:protocol}), not the
basis itself. Figure~\ref{fig:waveform_combined} summarizes the pulse construction:
panels (a),(e) compare the PSP against cosine DRAG; (b),(f) show the isolated ECP
tones; (c),(g) show the full PESP-S and PESP-C; and (d),(h) give the per-level
leakage trajectories.

The second-derivative term in Eq.~(\ref{eq:main}) uses the constant-subtracted
form
\begin{equation}
\tilde F_2(t)=\ddot F_\theta(t)-\ddot F_\theta(0)
=-\sum_n a_n\omega_n^2\, g_n(t),
\label{eq:reparam}
\end{equation}
which removes only the DC component of $\ddot F_\theta$ and therefore vanishes at
both endpoints for any parameters while remaining in the same cosine basis; the
AC content, and hence the leakage-suppression action near the
$\ket{1}$--$\ket{2}$ transition, is unchanged (endpoint residual $\lesssim10^{-15}$).

\subsection{Endpoint-cancellation pulse}

\subsubsection{First tone}

The endpoint-cancellation
pulse (ECP) is built from two tones, one near each leakage
transition. With the PSP fixed, the first tone, near the
$\ket{1}$--$\ket{2}$ transition, is
\begin{multline}
\Omega_{\mathrm{aux}}(t)=A_m r_{\mathrm{aux}}\,e^{i(\delta_{\mathrm{aux}}t+\phi)}\\
\qquad\times\bigg[\frac{i\dot F_\theta}{\delta_{\mathrm{aux}}}
+W(t)\bigg(b_2\frac{\ddot F_\theta}{\eta^2}
+i b_3\frac{F_\theta^{(3)}}{\eta^3}\bigg)\bigg].
\label{eq:aux}
\end{multline}
with $\delta_{\mathrm{aux}}=\delta_{\mathrm{ratio}}\,\eta\approx-\eta$ 
and the endpoint window $W(t)=F_\theta/\max|F_\theta|$. 
With the filter terms off ($b_2=b_3=0$) this tone reduces to the active leakage
cancellation (ALC) pulse~\cite{Chiaro2025}; activating $b_2$ and $b_3$ defines
the single-tone ECP. 
Their derivative/filter form follows the spectrally balanced construction of Ref.~\cite{Wang2025} 
and the broader multi-derivative DRAG strategy for suppressing multiple off-resonant transitions~\cite{Li2024}. 
The first-derivative term vanishes at the pulse boundaries by construction,
and the window $W(t)$ enforces the same boundary behavior for the $b_2$ and
$b_3$ terms. 
The $b_2$ term uses the raw second derivative, matching the stated analytic construction.

\subsubsection{Second tone}

In the rotating
frame of Eq.~(\ref{eq:H}) the level energies are $E_n=-\eta\,n(n-1)/2$, so the
$\ket{1}$--$\ket{2}$ gap is $E_2-E_1=-\eta$ (the transition the first tone
addresses) and the $\ket{2}$--$\ket{3}$ gap is $E_3-E_2=-2\eta$. The $\ket{3}$ component of the four-level endpoint floor
(Sec.~\ref{sec:secondtone}) arises from a cascade through the transiently
populated $\ket{2}$ state, with first-order endpoint amplitude
$c_3(T)\simeq-i\int_0^T\Lambda_{23}(t)\,c_2(t)\,e^{i2\eta t}\,dt$.
Motivated by this first-order amplitude, we suppress this contribution with a
second tone at carrier
$\delta_{\mathrm{aux2}}=\delta_{\mathrm{ratio2}}\,\eta\approx-2\eta$,
\begin{equation}
\Omega_{\mathrm{aux2}}(t)=A_m\,r_{\mathrm{aux2}}\,
e^{i(\delta_{\mathrm{aux2}}t+\phi_2)}\,W_2(t),
\label{eq:aux2}
\end{equation}
with the endpoint window $W_2(t)=F_\theta/\max|F_\theta|$, which vanishes
at both endpoints [$g_n(0)=g_n(T)=0$] and peaks where $\ket{2}$ is transiently
populated; the spectral weight added by this windowed tone is included in the
out-of-band fraction $C_{\mathrm{hf}}$ (Appendixes~\ref{app:stress} and
\ref{app:hardware}). We constrain the second-tone amplitude to
\(|r_{\mathrm{aux2}}|\le 0.06\).  At the operating point this corresponds to a peak scale
\(A_m |r_{\mathrm{aux2}}|/\eta \lesssim 10^{-2}\); the associated
second-order off-resonant shifts of the computational and
\(\ket{1}\)--\(\ket{2}\) transitions, of order
\((A_m r_{\mathrm{aux2}}/2\eta)^2\sim10^{-5}\text{--}10^{-4}\), are
corrected by the Stage~3 amplitude and virtual-$Z$ calibration. The three parameters
$\theta_{\mathrm{aux2}}=[r_{\mathrm{aux2}},\delta_{\mathrm{ratio2}},\phi_2]$ are
initialized by a linear-response cancellation of the four-level $\ket{3}$
endpoint amplitude and are off ($r_{\mathrm{aux2}}=0$) in the baseline design.
The isolated ECP tones are shown in panels (b),(f) of Fig.~\ref{fig:waveform_combined}.
The full PESP assembles these modules: PESP-S is the PSP with a single-tone ECP,
and PESP-C adds the second ECP tone; cosine DRAG is the reference
[panels (c),(g) of Fig.~\ref{fig:waveform_combined}].
The leakage trajectories for initial $\ket{0}$ and $\ket{1}$ appear in
panels (d),(h).

\begin{table*}[t]
\caption{\label{tab:params}Parameters of the PESP, their roles in the staged optimization, the spectral
or level-resolved quantity each acts on, and the associated recalibration
variable. The overall amplitude $A_m$ is determined by the gate-area constraint
and the final amplitude calibration and is not a free dimension.}
\begin{ruledtabular}
\begin{tabular}{llll}
parameter & stage & mechanism / controlled object & recalibration\\
\colrule
$a_2,a_3$       & 1 & cosine weights; path band, $\tilde\Lambda(\eta)$ at $\eta=\omega_2$ & pulse shape\\
$\beta_1$       & 1 & DRAG quadrature; leading $\ket{1}$--$\ket{2}$ coupling & ---\\
$\beta_2$       & 1 & second-order DRAG; residual $\ket{1}$--$\ket{2}$ & ---\\
$\delta_m$      & 1 & AC-Stark carrier detuning; $\Delta$ & ---\\
$r_{\mathrm{aux}},\delta_{\mathrm{ratio}},\phi$ & 2 & first ECP tone; $\ket{2}$ endpoint $\tilde\Lambda(\eta)$ & $\eta$-tracked detuning\\
$b_2,b_3$       & 2 & windowed filter terms; $\ket{2}$ endpoint (higher order) & ---\\
$r_{\mathrm{aux2}},\delta_{\mathrm{ratio2}},\phi_2$ & 2 & second ECP tone; $\ket{3}$ cascade endpoint & $2\eta$-tracked detuning\\
$A_m$           & 3 & overall amplitude; $\pi/2$ area & Rabi / virtual-$Z$\\
\end{tabular}
\end{ruledtabular}
\end{table*}

% ============================================================================
\section{Staged optimization protocol}\label{sec:protocol}
% ============================================================================

\subsection{Stage 1 and Stage 2 objectives}

We evaluate the six combinations formed from two main-pulse families
$\{\textrm{cosine DRAG},\textrm{PSP}\}$ and three endpoint corrections
$\{\textrm{none},\textrm{ALC pulse},\textrm{single-tone ECP}\}$; the
cosine-DRAG$\,+\,$single-tone-ECP configuration tests whether the auxiliary
filter terms reduce the transient exposure without path shaping. The two-tone
ECP is evaluated on the final PESP candidate (Stage-2 extension below). The Stage-1 and Stage-2 cost
functions are
\begin{align}
\mathcal{C}_{\mathrm{main}}&=w_I(1-F)+w_L\Pfinal+w_{\mathrm{path}}\Pbar
   \nonumber\\
&\qquad +w_M\max_t P_A+w_{\mathrm{hf}}C_{\mathrm{hf}},
   \label{eq:cmain}\\[4pt]
\mathcal{C}_{\mathrm{aux}}&=w_I(1-F)+w_L\Pfinal+w_{\mathrm{hf}}C_{\mathrm{hf}}
   \nonumber\\
&\qquad +\epsilon_{\mathrm{path}}\Pbar ,
   \label{eq:caux}
\end{align}
with $w_M=0.1w_{\mathrm{path}}$ and $\epsilon_{\mathrm{path}}=0$. 
Stage~2 therefore acts only on the endpoint. 
The choice $\epsilon_{\mathrm{path}}=0$ reflects the role of the ECP as a generalization of 
the ALC pulse: both suppress the endpoint amplitude using auxiliary tones near the leakage transitions, 
through a mechanism distinct from the transient-path suppression provided by the PSP. 
Adding a path penalty to the ECP optimization has negligible effect, 
producing a ${\lesssim}1\%$ change in $\Pbar$, 
because the auxiliary tones primarily cancel the endpoint residual 
and do not substantially reshape the drive envelope. 
The peak-leakage weight is tied heuristically to the path weight 
so that the integrated- and peak-exposure penalties co-scale, 
leaving the sweep effectively single-parameter. 
The operating point is insensitive to this prefactor: 
the path reduction changes by less than $1\%$ 
when $w_M/w_{\mathrm{path}}$ is swept from $0.01$ to $1$ (Appendix~\ref{app:numerics}).

\subsection{Stage 2 extension with a two-tone ECP}

Stages~1--2 as described above
optimize the first ECP tone at three levels. Because the level-resolved
four-level endpoint population contains distinct $\ket{2}$ and $\ket{3}$
contributions, associated with back-action and cascade processes
(Sec.~\ref{sec:sweep}), a second tone is required: with
the main pulse fixed, we jointly re-optimize the first tone
$\theta_{\mathrm{aux}}$ (in a trust region about its converged value) together
with the second-tone parameters $\theta_{\mathrm{aux2}}$
[Eq.~(\ref{eq:aux2})] against the four-level total endpoint
$\Pfinal=\Pfinal^{\ket{2}}+\Pfinal^{\ket{3}}$ plus the fidelity, followed by the
Stage~3 retuning. Optimizing the two tones jointly rather than sequentially allows the
optimization to account for their mutual influence: the first tone suppresses
the residual $\ket{2}$ endpoint population in the four-level calculation,
while the second tone suppresses the $\ket{3}$ contribution. This
configuration reaches the ${\sim}10^{-8}$ endpoint floor of
Sec.~\ref{sec:secondtone}.

\subsection{Stage 3 amplitude and virtual-$Z$ recalibration}

With Stages~1--2 fixed, we re-calibrate the overall amplitude
of the complete (main$+$auxiliary) pulse by a one-dimensional scan
(coarse $\pm5\%$/51 points, refined $\pm0.5\%$/31 points) that maximizes the
$Z$-optimized fidelity, and then record the explicit virtual-$Z$ phases. This
final one-dimensional calibration corrects the residual rotation introduced by
the endpoint tones; it leaves converged main-pulse designs essentially unchanged but
recovers fidelity for deliberately detuned designs.

\subsection{Numerical optimizer}

We optimize with CMA-ES (population $12$, $\sigma_0=0.2$).
Each trial is one equal-budget independent run, and we report the median over
five trials, restricting the median to the ``path basin'' where appropriate
(classification $P_{\rm path}^{3L}/P_{\rm path,cosine}^{3L}<0.95$). Stage~1 uses
a two-phase warm start (fidelity-only Phase~A, then full weights), with the
Phase-A budget carved out of the total so that warm and cold starts use the same
number of evaluations. An equal-budget comparison shows the warm start markedly
improves access to the path basin (warm $5/5$, cold $2/5$). A fully joint optimization is not used in the main analysis; the staged
procedure retains the parameter assignments introduced above, and a direct
four-level optimization is included as a consistency check in
Appendix~\ref{app:numerics}. The endpoint-to-infidelity weight ratios
$\rho_{\mathrm{main}}=w_L/w_I$ in Eq.~(\ref{eq:cmain}) and
$\rho_{\mathrm{aux}}=w_L/w_I$ in Eq.~(\ref{eq:caux}) form a broad plateau; we
adopt $(\rho_{\mathrm{main}},\rho_{\mathrm{aux}})=(0.2,7)$ as a representative
point.

% ============================================================================
\section{Four-level closed-system results}\label{sec:coherent}
% ============================================================================

We now report the closed-system results.  The three-level model enters only as a
low-dimensional generator of candidate pulses and as a diagnostic for the
endpoint-only optimum; the principal endpoint floors and the operating-point
selection are four-level quantities.  Unless stated otherwise, all entries in
this section are medians over five independent equal-budget CMA-ES seeds.

\subsection{Four-level operating-point selection}\label{sec:sweep}

\begin{figure}[t]
\centering
\includegraphics[width=\columnwidth]{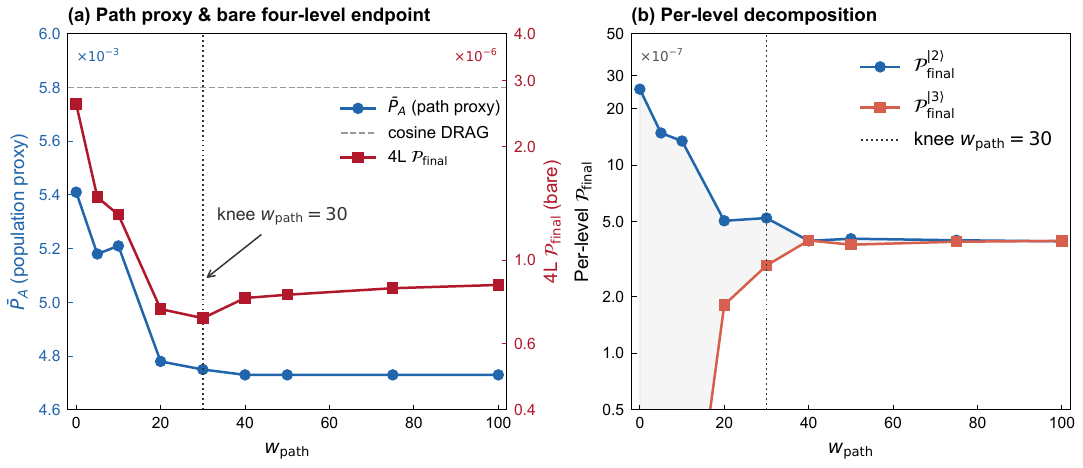}
\caption{\label{fig:knee}Four-level leakage-floor knee.
(a)~Path proxy $\Pbar$ (left axis, blue) and bare four-level $\Pfinal$
(right axis, red) versus $w_{\mathrm{path}}$;
basin occupancy shown at each point; the knee at $w_{\mathrm{path}}=30$
is marked.
(b)~Per-level decomposition into $\Ptwo$ ($\ket{2}$ back-action, falling)
and $\Pthree$ ($\ket{3}$ cascade, rising); the crossing coincides with the minimum of the total bare floor.}
\end{figure}

Figure~\ref{fig:knee} summarizes the path-weight
sweep for the PSP with a single-tone ECP.  The proxy path is the three-level
design quantity used by Stage~1, while the bare four-level endpoint floor is that
of the same pulse before the second ECP tone.  The four-level floor is
non-monotonic: it falls from $2.6\times10^{-6}$ at the endpoint-only point
$w_{\mathrm{path}}=0$ to $7.0\times10^{-7}$ near $w_{\mathrm{path}}=30$,
then rises toward $8.6\times10^{-7}$ as the path penalty is over-weighted.
This behavior is consistent with the first-order cascade amplitude
$c_3(T)\simeq-i\int_0^T\Lambda_{23}(t)\,c_2(t)\,e^{i2\eta t}\,dt$
(Sec.~\ref{sec:ansatz}): as $w_{\mathrm{path}}$ grows, the reshaping of
$c_2(t)$ that lowers its time average increases the net overlap feeding the
$\ket{3}$ channel, so $\Pthree$ rises while $\Ptwo$ falls, and the bare floor is
minimized where the two channels cross [Fig.~\ref{fig:knee}(b)].  We
therefore use $w_{\mathrm{path}}=30$ as the operating point: it lies at the
four-level leakage-floor knee and in the saturated path-reduction plateau.

The endpoint-only point illustrates the limitation of using an endpoint
zero as the final metric.  It reaches the three-level numerical floor but has a
larger path than cosine DRAG, occupies the path basin in only one of five seeds,
and has the largest four-level floor in the sweep.  The complete per-level
breakdown of the knee is given in Appendix~\ref{app:fourleveldecomp}.

\subsection{Operating-point mechanism matrix}\label{sec:matrix}

Table~\ref{tab:matrix} replaces the three-level mechanism matrix with the
corresponding four-level evaluation.  Comparisons between rows show the effect of changing the main pulse, and
comparisons between columns show the effect of adding an endpoint correction.
Changing from cosine DRAG to the PSP reduces the four-level dephasing
exposure $\Pbardeph$ from $6.24\times10^{-3}$ to $4.91\times10^{-3}$, a $21.3\%$ reduction, whereas
adding the ALC pulse or single-tone ECP at fixed main pulse mainly changes
endpoint quantities.
Seed-to-seed variation at the operating point is modest: the interquartile
spread of the path reduction across the five seeds is below three percentage
points, and four of the five seeds have bare four-level floors within ten
percent of the median, with the remaining seed converging to a shallower
optimum of the kind mitigated by the multi-seed restarts of
Sec.~\ref{sec:protocol}.
The table also shows why the second tone is needed: at the knee the bare
endpoint floor contains both a residual $\ket{2}$ contribution and a
non-negligible $\ket{3}$ component.

\begin{table*}[t]
\caption{\label{tab:matrix}Four-level operating-point mechanism matrix.  The endpoint columns
report the bare four-level floor before the second tone is added; the exposure
column is the four-level dephasing exposure $\Pbardeph$.  The row comparison
shows that the path reduction is produced by the PSP, while
the ALC pulse and single-tone ECP primarily act on endpoint amplitudes.
Each entry is a component-wise five-seed median, so the per-level entries
$\Ptwo$ and $\Pthree$ need not sum exactly to the median total $\Pfinal$.}
\begin{ruledtabular}
\begin{tabular}{lcccccc}
method & $1-F$ & 4L $\Pfinal$ & $\Ptwo$ & $\Pthree$ & 4L $\Pbardeph$ & $\Delta\Pbardeph$\\
\colrule
cosine DRAG                   & $4.21\times10^{-5}$ & $4.18\times10^{-5}$ & $4.18\times10^{-5}$ & $8.57\times10^{-10}$ & $6.24\times10^{-3}$ & ---\\
\quad$+$ALC                   & $6.81\times10^{-6}$ & $4.97\times10^{-6}$ & $4.97\times10^{-6}$ & $2.06\times10^{-9}$  & $6.04\times10^{-3}$ & $-3.3\%$\\
\quad$+$single-tone ECP       & $6.57\times10^{-6}$ & $5.14\times10^{-6}$ & $5.14\times10^{-6}$ & $1.87\times10^{-9}$  & $6.03\times10^{-3}$ & $-3.3\%$\\
path-shaping pulse            & $8.31\times10^{-7}$ & $7.71\times10^{-7}$ & $5.23\times10^{-7}$ & $2.92\times10^{-7}$  & $4.91\times10^{-3}$ & $-21.3\%$\\
\quad$+$ALC                   & $7.96\times10^{-7}$ & $7.29\times10^{-7}$ & $4.34\times10^{-7}$ & $2.92\times10^{-7}$  & $4.91\times10^{-3}$ & $-21.3\%$\\
\quad$+$single-tone ECP       & $7.62\times10^{-7}$ & $7.01\times10^{-7}$ & $4.33\times10^{-7}$ & $2.53\times10^{-7}$  & $4.91\times10^{-3}$ & $-21.3\%$\\
\end{tabular}
\end{ruledtabular}
\end{table*}

\subsection{Matched-budget spectral-null baselines}\label{sec:fastbaseline}

Two matched-budget controls test whether the path reduction is reachable by
endpoint or spectral-null shaping. The first is an endpoint-only optimization of
our own ansatz ($w_{\mathrm{path}}=0$), which drives the endpoint to its
minimum within the same low-dimensional family. The second is an analytic
FAST~DRAG spectral-null pulse~\cite{Hyyppa2024}, constructed at the same gate budget by minimizing the in-phase spectral energy
over a band around the $\ket{1}$--$\ket{2}$ transition
(Appendix~\ref{app:fast}). Table~\ref{tab:fast}
collects both alongside the path-shaped and two-tone designs.

The endpoint-only optimization reduces the four-level endpoint leakage by
more than an order of magnitude relative to cosine DRAG while changing
$\Pbardeph$ by only $+0.6\%$. FAST~DRAG-L lowers the endpoint floor by a
factor of approximately three, to $1.4\times10^{-5}$, and changes $\Pbardeph$
by $-2.4\%$, an order of magnitude less than the $-21.3\%$ of the PSP; the
Lindblad excess follows the same pattern ($-1.6\%$ versus $-20.4\%$,
Sec.~\ref{sec:dephasing}). The endpoint values of the two constructions are
not directly comparable: FAST~DRAG optimizes a spectral notch rather than the
absolute endpoint leakage, and a fully calibrated FAST or HD~DRAG
implementation can reach lower endpoint floors (Appendix~\ref{app:fast}). The
comparison relevant here is the transient exposure, for which neither
endpoint-only nor spectral-notch optimization produces a reduction comparable
to that of the path-shaping pulse. This is consistent with the analysis of
Sec.~\ref{sec:mechanism}: a control that nulls $\tilde\Lambda(\eta)$
constrains the endpoint sample but not the surrounding band that sets
$\Pbar$. The two-tone ECP subsequently lowers the endpoint floor while leaving
$\Pbardeph$ unchanged at the reported precision.

\begin{table*}[t]
\caption{\label{tab:fast}Four-level matched-budget comparison. The endpoint-only row uses the same
ansatz and optimization budget with $w_{\mathrm{path}}=0$; the FAST~DRAG-L row
is an analytic spectral-null pulse~\cite{Hyyppa2024} constructed at the same
gate duration (Appendix~\ref{app:fast}). Endpoint leakage and transient
exposure are reported separately because the pulse families optimize different
objectives.}
\begin{ruledtabular}
\begin{tabular}{llcccc}
method & objective & $1-F$ & 4L $\Pfinal$ & 4L $\Pbardeph$ & $\Delta\Pbardeph$\\
\colrule
cosine DRAG & calibrated baseline & $4.21\times10^{-5}$ & $4.18\times10^{-5}$ & $6.24\times10^{-3}$ & ---\\
FAST DRAG-L & spectral notch~\cite{Hyyppa2024} & $1.50\times10^{-5}$ & $1.43\times10^{-5}$ & $6.09\times10^{-3}$ & $-2.4\%$\\
PSP ansatz & $w_{\mathrm{path}}=0$, endpoint-only & $2.86\times10^{-6}$ & $2.68\times10^{-6}$ & $6.28\times10^{-3}$ & $+0.6\%$\\
PSP$+$single-tone ECP & $w_{\mathrm{path}}=30$, knee point & $7.62\times10^{-7}$ & $7.01\times10^{-7}$ & $4.91\times10^{-3}$ & $-21.3\%$\\
PSP$+$two-tone ECP & four-level endpoint optimization & $2.96\times10^{-7}$ & $2.96\times10^{-8}$ & $4.91\times10^{-3}$ & $-21.3\%$\\
\end{tabular}
\end{ruledtabular}
\end{table*}

\subsection{Two-tone endpoint correction}\label{sec:secondtone}

The per-level decomposition identifies two addressable endpoint channels.  The
first ECP tone is re-optimized against the four-level $\ket{2}$ endpoint, and the
second tone near $-2\eta$ targets the $\ket{3}$ population associated with
the cascade process.  The joint two-tone optimization followed by the Stage-3
calibration lowers the endpoint floor by more than an order of magnitude while
improving the projected fidelity.  At the
operating point (Table~\ref{tab:secondtone}; the full knee-region sweep is
given in Appendix~\ref{app:fourleveldecomp}) the single-tone ECP falls from $7.01\times10^{-7}$
to $2.96\times10^{-8}$ in $\Pfinal$, and the residual is almost entirely the
$\ket{3}$ component.  Although $w_{\mathrm{path}}=20$ gives the smallest two-tone endpoint floor
in this sweep, we retain $w_{\mathrm{path}}=30$ as the operating point: it
minimizes the bare four-level floor across the sweep and lies in the saturated
path-reduction plateau (Fig.~\ref{fig:knee}), a criterion that uses only
quantities available before the second-tone stage.

\begin{table*}[t]
\caption{\label{tab:secondtone}Four-level endpoint floor at the operating point
before and after the second ECP tone.  The two-tone ECP suppresses the
$\ket{2}$ endpoint contribution to the $10^{-9}$ level and leaves a
predominantly $\ket{3}$ residual.  The full knee-region sweep is given in
Appendix~\ref{app:fourleveldecomp}.
Entries are component-wise five-seed medians; per-level values need not sum
exactly to $\Pfinal$.}
\begin{ruledtabular}
\begin{tabular}{lcccc}
config. & $1-F$ & $\Pfinal$ & $\Ptwo$ & $\Pthree$\\
\colrule
PSP$+$ALC       & $8.0\times10^{-7}$ & $7.29\times10^{-7}$ & $4.34\times10^{-7}$ & $2.92\times10^{-7}$\\
PESP-S  & $7.6\times10^{-7}$ & $7.01\times10^{-7}$ & $4.33\times10^{-7}$ & $2.53\times10^{-7}$\\
PSP$+$ALC$+$second ECP tone & $5.0\times10^{-7}$ & $1.85\times10^{-7}$ & $5.1\times10^{-9}$  & $1.50\times10^{-7}$\\
PESP-C          & $3.0\times10^{-7}$ & $2.96\times10^{-8}$ & $1.4\times10^{-9}$  & $2.79\times10^{-8}$\\
\end{tabular}
\end{ruledtabular}
\end{table*}

A natural concern is that suppressing $\ket{3}$ might shift leakage upward into
$\ket{4}$.  A five-level evaluation of the PESP-C design yields $\Pfinal^{\ket{4}}\sim
6.5\times10^{-12}$ (median), less than $0.03\%$ of the total leakage and three
orders of magnitude below the $\ket{3}$ residual.  For this operating point,
the five-level calculation supports the adequacy of the four-level truncation
(Appendix~\ref{app:fivelevel}).

% ============================================================================
\section{Four-level Lindblad dephasing validation}\label{sec:dephasing}
% ============================================================================

We next test whether the reduction in the closed-system exposure proxy is
reflected in the leakage obtained from noisy four-level dynamics, using
$\Pexc$ as an output of the simulation rather than as an optimized cost
term.  We use a four-level pure-dephasing Lindblad propagator with
$L=n$, $\gamma_\phi=2/\Tphi$, and $\Tphi\in\{5,10,20,50,100,200,500,1000,2000,5000\}~\mu$s.  The noisy
leakage is evaluated at the four-level operating point and compared with the
first-order prediction $\Pexc\simeq\gamma_\phi T\Pbardeph$.

\begin{figure}[t]
\centering
\includegraphics[width=\columnwidth]{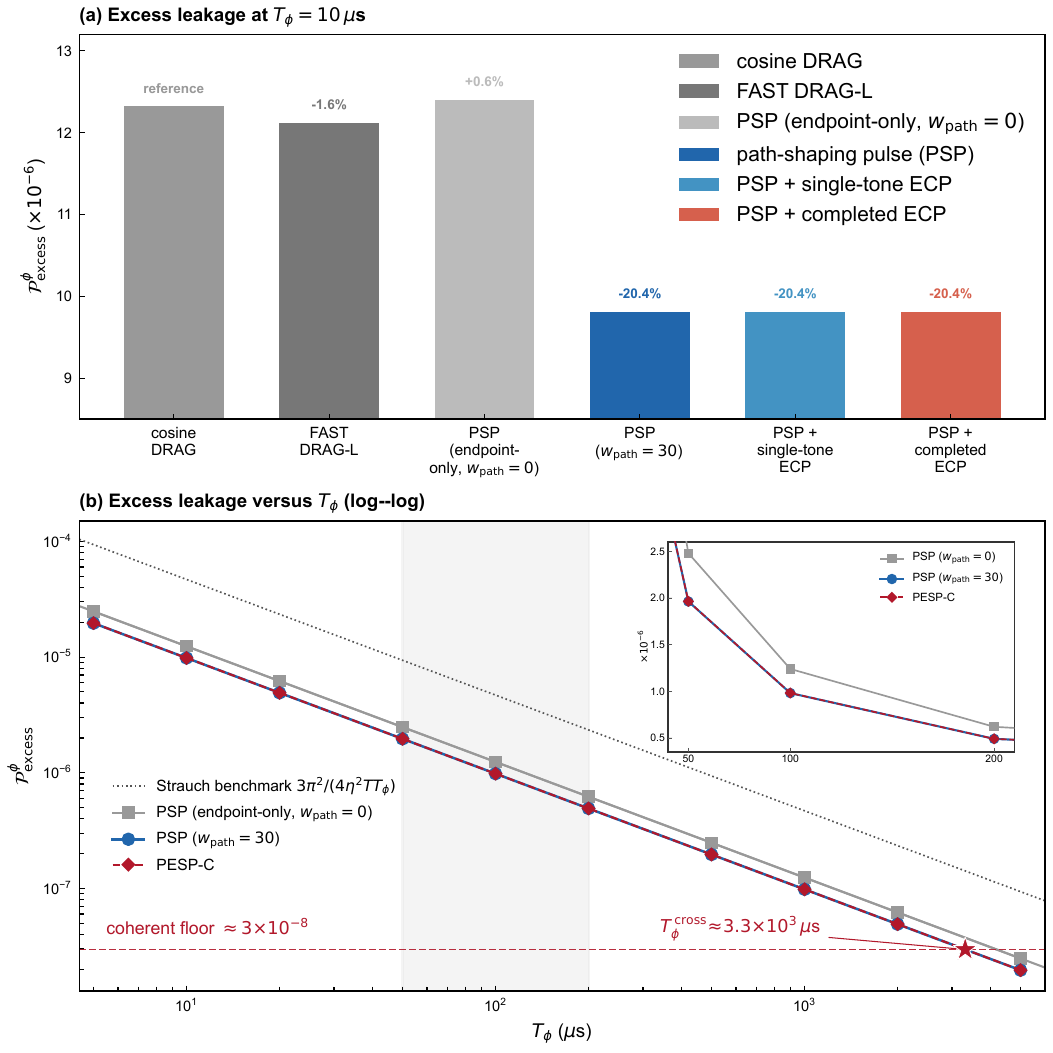}
\caption{\label{fig:dephasing}Four-level Lindblad pure-dephasing validation and
staged pulse construction.
(a) Four-level pure-dephasing excess leakage at $\Tphi=10~\mu$s, normalized by
the cosine-DRAG value.  The endpoint-only and FAST-class baselines remain near
the cosine value, whereas the path-shaping pulse reduces the excess by about
$20\%$.  Adding the single-tone or two-tone ECP leaves this reduction unchanged,
consistent with the ECP acting on endpoint amplitudes rather than on the
transient exposure.
(b) Schematic of the staged PESP construction.  The main pulse controls the
transient leakage path, the first ECP tone suppresses the residual $\ket{2}$ endpoint amplitude,
and the second tone suppresses the $\ket{3}$ cascade contribution.}
\end{figure}

Figure~\ref{fig:dephasing}(a) shows the $\Tphi=10~\mu$s representative point.  The
endpoint-only design again behaves like cosine DRAG: it has no path advantage and
therefore no reduction in dephasing-induced excess.  In contrast, the path-aware
designs reduce the excess by about $20\%$, and the exposure inferred from
$\Pexc/(\gamma_\phi T)$ agrees with the closed-system four-level
$\Pbardeph$.  Adding the single-tone or two-tone ECP leaves $\Pbardeph$
unchanged at the reported precision, showing that the ECP tones act on
endpoint amplitudes rather than on the transient exposure.

For orientation, the white-noise benchmark of Strauch,
$3\pi^2/(4\eta^2 T\Tphi)=4.69\times10^{-5}$ at $\Tphi=10~\mu$s,
sets the relevant order of magnitude for dephasing-induced leakage at the
present $\eta$ and $T$. This expression is not a pulse-specific prediction for
our shaped $\pi/2$ gates: it is derived as a white-noise benchmark for a
specific single-qubit gate envelope. Our simulated excess leakage is smaller by
a factor of about $4$--$5$, consistent with the smaller gate area and pulse
shaping, and therefore with the benchmark as a scale estimate rather than an
exact equality.

\subsection{Long-dephasing-time behavior}

Over the simulated range the
excess leakage follows the expected $1/\Tphi$ scaling
(Fig.~\ref{fig:dephasing}), so the relative reduction provided by the
path-aware design, about $21\%$ in $\Pbardeph$ and $20\%$ in $\Pexc$, is
approximately independent of $\Tphi$ over this range. 
The absolute excess decreases as $\Tphi$ grows, 
raising the question of whether the reduction remains meaningful once $\Pexc$ is small. 
Two considerations show that it does. First, the relevant
comparison is not against zero but against the coherent endpoint floor after the two-tone correction,
$\Pfinal\approx3\times10^{-8}$. With $C\equiv\Pexc\,\Tphi\approx9.8\times10^{-5}~
\mu$s, the path-aware excess exceeds this floor until
$\Tphi\approx C/\Pfinal\simeq3.3\times10^{3}~\mu$s, a crossover bracketed
directly by the simulated grid, whose $\Tphi=2000$ and $5000~\mu$s points lie
just above and below the floor (Fig.~\ref{fig:dephasing}). For dephasing times up to several hundred microseconds, which covers
reported transmon coherence times~\cite{Kjaergaard2020}, the dephasing channel
rather than the coherent floor sets the residual leakage. Only beyond
$\Tphi\sim3\times10^{3}~\mu$s, well above this range, does the coherent
floor dominate and the reduction in total leakage saturate. Second, because leakage falls outside the Pauli error model and can be
converted by subsequent entangling gates into correlated
errors~\cite{McEwen2021,Miao2023}, reducing transient exposure remains useful
even when the absolute excess is small. Finally,
although the single-qubit gate studied here provides a minimal setting for
isolating
the path--endpoint distinction, the same mechanism (dephasing acting on
transient auxiliary-state population) also appears in two-qubit and more
general multilevel operations. Ref.~\cite{Strauch2025} reports
dephasing-induced leakage in the $10^{-6}$--$10^{-4}$ range for experimentally
relevant single- and two-qubit gates, indicating that the transient channel can
be more consequential in larger multilevel operations.

\subsection{Spectrum-tailored generalization}

The population proxy and its
dephasing counterpart used above describe a \emph{white-noise} dephasing
environment; they are the $S(\omega)=S_0$ special case of the general
first-order leakage functional of Ref.~\cite{Strauch2025},
\begin{multline}
P_{\mathrm{leak}}=\frac{1}{\dQ}\sum_{j,k}\int_{-\infty}^{\infty}
\frac{d\omega}{2\pi}\,S(\omega)\,\big|\tilde A_{j\to k}(\omega)\big|^{2},\\
\tilde A_{j\to k}(\omega)=\int_0^T\!e^{i\omega t}A_{j\to k}(t)\,dt ,
\label{eq:general_spectrum}
\end{multline}
with $S(\omega)$ the dephasing power spectral density; for white noise
$S(\omega)=S_0=2/\Tphi$ Parseval's theorem reduces
Eq.~(\ref{eq:general_spectrum}) to $\gamma_\phi T\Pbardeph$, recovering
Eq.~(\ref{eq:firstorder}). We adopt white noise as a transparent,
device-independent demonstration, but the construction is not tied to it: given
a measured spectrum $S(\omega)$, one may instead minimize the spectrally
weighted exposure $\int(d\omega/2\pi)\,S(\omega)|\tilde A(\omega)|^2$, shaping
the running excursion where the noise actually carries weight, in the spirit of
filter-function engineering~\cite{Cywinski2008,Green2013,PazSilva2014,Oda2023}. Because
low-frequency ($1/f$) noise concentrates its weight at frequencies where
$|\tilde A(\omega)|^2$ is suppressed (Ref.~\cite{Strauch2025} accordingly finds $1/f$ leakage smaller
than the white-noise value at equal dephasing time), the white-noise
functional provides a useful reference case, while a measured
noise spectrum could enable further improvement through spectrum-aware shaping. A full spectrum-tailored
optimization is left to future work.

% ============================================================================
\section{Discussion}\label{sec:discussion}
% ============================================================================

\subsection{Scope and stress boundary}

The main results are obtained at
$T=10$~ns and $\eta/2\pi=0.200$~GHz. Across the tested grid
$T\in\{6,8,10,12,15\}$~ns and
$\eta/2\pi\in\{0.15,\dots,0.25\}$~GHz, the PSP yields an
${\sim}11$--$20\%$ reduction in transient exposure relative to cosine DRAG at
the points for which the optimizer converges to the path basin
(Appendix~\ref{app:stress}). The optimized parameters are not directly
transferable between gate times and anharmonicities: $w_{\mathrm{path}}$ must be
reselected and the ECP tones recalibrated. At the shortest-gate,
smallest-anharmonicity point ($6$~ns, $0.15$~GHz), the two-tone pulse no longer
satisfies the waveform constraints adopted here. The calibration tests in
Appendix~\ref{app:hardware} further show that the Stage-1 exposure reduction is
relatively insensitive to modest parameter errors, whereas the endpoint floor
is sensitive to the calibration of the auxiliary tones, particularly the tone
near $-2\eta$.

The results distinguish a spectral endpoint zero, the residual endpoint
leakage in the enlarged Hilbert space, and the transient exposure
$\Pbardeph$. The first is a spectral diagnostic, the second quantifies the
coherent leakage remaining after the pulse, and the third determines the
leading dephasing-induced contribution. Endpoint optimization can reduce the
second quantity without appreciably reducing the third, as shown by the
matched-budget endpoint-only and FAST~DRAG baselines, both of which leave the
transient exposure within a few percent of the cosine-DRAG value
(Sec.~\ref{sec:coherent}). The filter terms likewise do not directly
reduce transient exposure. Their benefit appears after the two endpoint tones
are jointly optimized: among the pulse families examined, they produce a more
favorable residual amplitude for the second-tone correction and yield the
lowest endpoint floor after the two-tone correction. This improvement should therefore be
attributed to endpoint correction rather than to transient-path shaping.

Several limitations remain. The calculations are open loop and use an idealized
Hamiltonian and noise description. They do not include hardware closed-loop
calibration, transfer-function identification, two-level-system defects,
thermal excitation, or finite-bit-depth waveform generation. In particular, the
active leakage cancellation tone is treated throughout as an endpoint
correction, and the coherent endpoint floor of approximately $3\times10^{-8}$
is a closed-system quantity: it is not directly comparable to the hardware
leakage below $10^{-5}$ reported in Ref.~\cite{Chiaro2025}, where incoherent
heating sets a substantially higher background. A direct
four-level optimization for one representative seed provides a consistency
check and reaches the same qualitative regime
(Appendix~\ref{app:numerics}). However, a systematic many-seed four-level
optimization has not been performed. The reported endpoint floors therefore
characterize the present compact construction rather than the best
values attainable by unrestricted four-level optimal control.

\subsection{Relation to prior work}

The transient functional $\Pbardeph$
[Eq.~(\ref{eq:deph})] is motivated by the first-order dephasing treatment of
Strauch~\cite{Strauch2025}. Here it is used as a design proxy rather than as a
strictly equivalent functional. Related multiobjective approaches have treated
transient leakage as an independent optimization target. Poggi and
Kiely~\cite{Poggi2025} sequentially minimize gate error and either robustness
or time-integrated leakage, revealing a trade-off between the latter two
objectives. McCord, Kuzmanovi\'c, and Paraoanu~\cite{McCord2025} use a Pareto
objective that penalizes the maximum transient population of the second excited
state together with sensitivity to detuning.

The present work instead considers the relation between transient exposure and
residual endpoint leakage. The transient functional is connected to a specific
noise-induced observable through the first-order relation
$\Pexc\simeq\gamma_\phi T\Pbardeph$ and is tested independently using Lindblad
simulations in which $\Pexc$ is not part of the optimization objective
(Sec.~\ref{sec:dephasing}). The four-level calculation further resolves the
residual endpoint leakage into contributions associated with the $\ket{2}$
back-action and the $\ket{3}$ cascade, which motivate the two-tone endpoint
correction. How this path--endpoint trade-off interacts with robustness to
control errors remains an open question.

% ============================================================================
\section{Conclusion}\label{sec:conclusion}
% ============================================================================

We have numerically investigated transient exposure and residual endpoint leakage as
separate control objectives for fast single-qubit gates in weakly anharmonic
transmons. For a $10$~ns $R_X(\pi/2)$ gate with
$\eta/2\pi=0.2$~GHz, the path-shaping pulse reduces the dephasing exposure
by $21.3\%$ relative to cosine DRAG and reduces the excess leakage obtained from
independent Lindblad simulations by approximately $20\%$. By contrast, the
matched-budget endpoint-only control produces no comparable reduction in
transient exposure. These results show that a small closed-system endpoint
leakage does not by itself imply reduced leakage in the presence of dephasing.

In the four-level calculation, the residual endpoint leakage after path shaping
contains contributions associated with the $\ket{2}$ back-action and the
$\ket{3}$ cascade. A two-tone endpoint correction suppresses these
contributions and lowers the coherent endpoint leakage from approximately
$7\times10^{-7}$ to $3\times10^{-8}$ at the selected operating point, while
leaving the transient exposure essentially unchanged. Within the parameter
regime studied here, transient path shaping and endpoint correction therefore
address different components of the leakage error. The extent to which this separation persists under measured noise spectra,
waveform distortions, and closed-loop hardware calibration remains to be
established experimentally.

\begin{acknowledgments}
The authors thank Haoran He and the members of Zheng Shan's group at
Information Engineering University for helpful discussions.
\end{acknowledgments}

\section*{Data availability}
The data that support the findings of this article are available from the
authors upon reasonable request.

\clearpage
\appendix

% ============================================================================
\section{Three-level proxy diagnostics}\label{app:proxy}
% ============================================================================
\suppressfloats[t]

The three-level truncation is used to generate low-dimensional candidate pulses
and to diagnose endpoint-only optima.  It is not used as the final physical
leakage claim in the main text.  Table~\ref{tab:app_matrix3l} gives the original
three-level mechanism matrix at the operating point.  The corresponding
four-level matched-budget comparison is reported in the main text
(Table~\ref{tab:fast}): the endpoint-only baseline ($w_{\mathrm{path}}=0$)
improves $\Pfinal$ over cosine DRAG but leaves the path slightly above it
($+0.6\%$), whereas the path-aware design reduces it by $21.3\%$ and the two-tone
ECP further lowers $\Pfinal$ to $2.96\times10^{-8}$.  The three-level ``numerical
floor'' at $w_{\mathrm{path}}=0$ does not survive at four levels.

\begin{table}[!h]
\caption{\label{tab:app_matrix3l}Three-level proxy operating-point mechanism
matrix ($w_{\mathrm{path}}=30$, median over five seeds; path rows are
path-basin median, $5/5$). Endpoint values at the numerical floor are written
``num.\ floor.''}
\begin{ruledtabular}
\begin{tabular}{lccc}
method & $1-F$ & $\Pfinal$ & $\Pbar$ \\
\colrule
cosine DRAG             & $2.38\times10^{-5}$ & $2.36\times10^{-5}$ & $5.80\times10^{-3}$\\
\quad$+$single-tone ECP & $3.15\times10^{-6}$ & $8.7\times10^{-9}$  & $5.69\times10^{-3}$\\
\quad$+$two-tone ECP    & $2.60\times10^{-6}$ & $1.3\times10^{-8}$  & $5.69\times10^{-3}$\\
path-shaping pulse & $2.0\times10^{-9}$  & $1.6\times10^{-9}$  & $4.75\times10^{-3}$\\
\quad$+$single-tone ECP & $1.2\times10^{-9}$  & $1.4\times10^{-11}$ & $4.75\times10^{-3}$\\
\quad$+$two-tone ECP    & num.\ floor         & num.\ floor         & $4.75\times10^{-3}$\\
\end{tabular}
\end{ruledtabular}
\end{table}

Throughout the main text the population proxy
$\Pbar=\frac{1}{T}\int_0^T|c_2(t)|^2dt$ [Eq.~(\ref{eq:path})] is used for
the path-weight sweep, while the dephasing proxy
$\Pbardeph=\frac{1}{T\dQ}\sum_{j,k}\int_0^T|A_{j\to k}(t)|^2dt$ [Eq.~(\ref{eq:deph})] is
used for the pure-dephasing validation.  The two proxies are numerically nearly
identical in the Pareto region.
Table~\ref{tab:app_proxy_concordance} reports the five-seed median of each
proxy for the bare PSP across the knee region
($w_{\mathrm{path}}\ge 30$, $n_{\mathrm{steps}}=2400$).  The ratio
$\Pbar/\Pbardeph$ is $0.992\pm0.001$, and the worst-case single-seed
deviation from unity is $1.0\%$; the proxies are therefore interchangeable
for the purpose of path-weight selection and Pareto analysis.

\begin{table}[!h]
\caption{\label{tab:app_proxy_concordance}Population proxy and dephasing proxy
for PSP, five-seed medians ($n_{\mathrm{steps}}=2400$).}
\begin{ruledtabular}
\begin{tabular}{lcccc}
$w_{\mathrm{path}}$ & $\Pbar$ ($10^{-3}$) & $\Pbardeph$ ($10^{-3}$) & $\Pbar/\Pbardeph$ & max deviation\\
\colrule
30  & 4.751 & 4.793 & 0.991 & $1.0\%$\\
50  & 4.730 & 4.770 & 0.992 & $0.9\%$\\
75  & 4.726 & 4.766 & 0.992 & $0.8\%$\\
100 & 4.725 & 4.765 & 0.992 & $0.8\%$\\
\end{tabular}
\end{ruledtabular}
\end{table}

\FloatBarrier

% ============================================================================
\section{Spectral representation of the path functional}\label{app:fejer}
% ============================================================================

This appendix derives the exact spectral representation quoted in
Eq.~(\ref{eq:fejer}).  With $h(t)\equiv\Lambda(t)e^{i\eta t}$, the first-order
path functional is the quadratic form
$T^{-1}\!\int_0^T|c_2|^2dt=\int_0^T\!\!\int_0^T h(t)h^*(t')\,M(t,t')\,dt\,dt'$
with kernel $M(t,t')=1-\max(t,t')/T$ [Eq.~(\ref{eq:path_kernel})].  Writing
$\max(t,t')=[t+t'+|t-t'|]/2$ splits the kernel exactly into a stationary and a
non-stationary part,
\begin{equation}
M(t,t')=\frac{1}{2}\Big[1-\frac{|t-t'|}{T}\Big]
+\frac{1}{2}\Big[1-\frac{t+t'}{T}\Big].
\label{eq:kernel_split}
\end{equation}
The stationary (triangle) part depends only on $t-t'$ and diagonalizes in the
Fourier basis: its transform is the Fej\'er weight,
$\int_{-T}^{T}(1-|\tau|/T)\,e^{i\nu\tau}d\tau=\mathcal{F}_T(\nu)$, which yields
the band integral of Eq.~(\ref{eq:fejer}) with prefactor $1/2$.  The
non-stationary part depends only on the mean time $(t+t')/2$ and factorizes:
using $1-(t+t')/T=(1-t/T)+(1-t'/T)-1$,
\begin{multline}
\mathcal{R}=\operatorname{Re}\!\big[\tilde\Lambda(\eta)\,B^{*}\big]
-\tfrac{1}{2}\big|\tilde\Lambda(\eta)\big|^{2},\\
B=\int_0^T\Big(1-\frac{t}{T}\Big)\Lambda(t)\,e^{i\eta t}\,dt ,
\label{eq:fejer_remainder}
\end{multline}
since $\int_0^T h\,dt=\tilde\Lambda(\eta)$ is the endpoint sample of
Eq.~(\ref{eq:final_spectral}).  The remainder is therefore linear in the
endpoint amplitude, bounded by
$|\mathcal{R}|\le|\tilde\Lambda(\eta)|\,[\,|B|+|\tilde\Lambda(\eta)|/2\,]$, and
vanishes identically for any pulse satisfying the spectral-null condition
$\tilde\Lambda(\eta)=0$.  
For the calibrated pulses considered in this work the endpoint amplitude is
suppressed by design, so the Fej\'er band integral carries the path
functional. The representation is exact but serves interpretation only: in
all optimizations and tables the path is evaluated directly in the time
domain [Eqs.~(\ref{eq:path}) and (\ref{eq:deph})], so no reported value
depends on evaluating Eq.~(\ref{eq:fejer}) numerically.
\FloatBarrier

% ============================================================================
\section{Construction of the FAST DRAG-L baseline}\label{app:fast}
% ============================================================================

The matched-budget FAST~DRAG-L control of Sec.~\ref{sec:fastbaseline} is built
analytically, with no CMA-ES optimization, following the spectral-minimization
construction of Ref.~\cite{Hyyppa2024}.  The in-phase envelope is expanded in
the same cosine basis as the PSP, $g_n(t)=1-\cos(2\pi n t/T)$ with $N=4$
terms.  The coefficients minimize the in-phase spectral energy over the
stopband $[0.95,1.05]\times\eta/2\pi$ centered on the
$\ket{1}$--$\ket{2}$ transition, normalized by the in-band spectral energy up
to the cutoff $2\eta/2\pi$; the constrained quadratic problem is solved as a
generalized eigenvalue problem, and the gate area fixes the overall amplitude.
The leakage-minimizing DRAG quadrature ($\beta=1$) supplies the imaginary
component, and the pulse then receives the same amplitude calibration and
virtual-$Z$ extraction as every other method in the comparison.  The stopband
half-width ($5\%$ of the transition frequency) and cutoff follow the heuristics
of Ref.~\cite{Hyyppa2024} and are not tuned against our cost functions; a fully
calibrated FAST/HD~DRAG implementation with device-specific hyperparameter
selection can reach a lower endpoint floor than the value quoted in
Table~\ref{tab:fast}; endpoint leakage and transient exposure are therefore
reported separately there.

\FloatBarrier

% ============================================================================
\section{Level truncation and numerical validation}\label{app:levelnumerics}
% ============================================================================

This appendix collects the validation checks that establish the stability of the
four-level operating-point claims.  We first decompose the four-level endpoint
floor across the knee, then verify that completing the endpoint-cancellation
pulse does not move leakage into a fifth level, and finally check time-step
convergence, sensitivity to the peak-leakage weight, and one direct four-level
optimization seed.

\subsection{Four-level endpoint floor and per-level decomposition}\label{app:fourleveldecomp}
\suppressfloats[t]

The sweep over $w_{\mathrm{path}}$ reveals a non-monotonic four-level endpoint floor. 
The decrease in $\ket{2}$ back-action and the increase in the $\ket{3}$ 
cascade explain the four-level knee used in the main text. 
Across this knee, the bare floor is nearly independent of the endpoint module: 
the main pulse, the single-tone ECP, and the two-tone ECP agree to within 
${\sim}10\%$ (cf.\ Table~\ref{tab:matrix}). 
This indicates that the floor is set by the main pulse rather than by the endpoint tones. 
Table~\ref{tab:app_secondtone} gives the knee-region floor before and after the second ECP tone, 
with the operating-point values summarized in Table~\ref{tab:secondtone}.

\begin{table}[!b]
\caption{\label{tab:app_secondtone}
Knee-region endpoint floor before and after the second ECP tone
(median over five seeds, $n_{\mathrm{steps}}=2400$).  The bare column is the
single-tone ECP before the second tone is added.  Across the knee region, the
two-tone correction lowers the endpoint floor by more than one order of
magnitude, suppresses the $\ket{2}$ endpoint contribution to the $10^{-9}$
level, and leaves a
predominantly $\ket{3}$ residual.  Entries are component-wise five-seed
medians; per-level values need not sum exactly to the total.}
\begin{ruledtabular}
\begin{tabular}{rcccc}
$w_{\mathrm{path}}$
& bare $\Pfinal$
& comp.\ $\Pfinal$
& comp.\ $\Ptwo$
& comp.\ $\Pthree$\\
\colrule
20 & $7.41{\times}10^{-7}$ & $1.66{\times}10^{-8}$ & $5.0{\times}10^{-10}$ & $1.61{\times}10^{-8}$\\
30 & $7.01{\times}10^{-7}$ & $2.96{\times}10^{-8}$ & $1.4{\times}10^{-9}$  & $2.79{\times}10^{-8}$\\
40 & $8.10{\times}10^{-7}$ & $7.6{\times}10^{-8}$  & $1.2{\times}10^{-9}$  & $7.4{\times}10^{-8}$\\
50 & $8.09{\times}10^{-7}$ & $7.78{\times}10^{-8}$ & $1.3{\times}10^{-9}$  & $7.65{\times}10^{-8}$\\
\end{tabular}
\end{ruledtabular}
\end{table}

\subsection{Five-level truncation verification}\label{app:fivelevel}
\suppressfloats[t]

\begin{figure}[!h]
\centering
\includegraphics[width=\columnwidth]{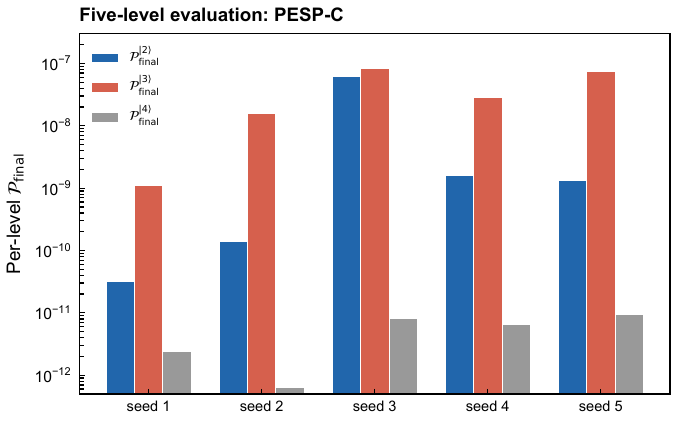}
\caption{\label{fig:fivelevel}Five-level per-level decomposition of the PESP-C design at
$w_{\mathrm{path}}=30$, evaluated seed by seed.
$\Pfinal^{\ket{2}}$, $\Pfinal^{\ket{3}}$, and $\Pfinal^{\ket{4}}$ on a log
scale.  The $\ket{4}$ component is three to four orders of magnitude below the
dominant $\ket{3}$ channel and negligible.}
\end{figure}

To check that the second ECP tone does not merely push the leakage cascade
from $\ket{3}$ into $\ket{4}$, we evaluated the PESP-C design at the operating point
$w_{\mathrm{path}}=30$ with a five-level model
($n_{\mathrm{steps}}=2400$).  Figure~\ref{fig:fivelevel} reports the per-level
decomposition of the five-level endpoint floor, seed by seed.  The $\ket{4}$
population is $\sim 6.5\times10^{-12}$ at the median, corresponding to a
fraction of $2.2\times10^{-4}$ of the total leakage, and never exceeds $0.3\%$
in any seed.  The $\ket{4}$
channel is therefore more than three orders of magnitude below the $\ket{3}$
residual and negligible.  For this operating point the four-level truncation is
adequate.

\FloatBarrier

\subsection{Numerical convergence, peak-leakage sensitivity, and direct four-level sanity check}\label{app:numerics}
\suppressfloats[t]

The two-tone operating-point result is stable against time-step
refinement.  Table~\ref{tab:app_convergence} reports the five-seed median at
$n_{\mathrm{steps}}=1200,2400,4800$.  The endpoint floor changes by less
than $0.1\%$ from $2400$ to $4800$ steps.  As a separate sanity check,
Table~\ref{tab:app_true4l} reports one direct four-level optimization seed.  It
reaches the same qualitative regime, but this single-seed result is not used as a
main-text claim.

\begin{table}[!b]
\caption{\label{tab:app_convergence}Time-step convergence for the PESP-C
design at $w_{\mathrm{path}}=30$
(five-seed median).}
\begin{ruledtabular}
\begin{tabular}{lccc}
$n_{\mathrm{steps}}$ & $1-F$ & 4L $\Pfinal$ & 4L $\Pbardeph$\\
\colrule
1200 & $2.966\times10^{-7}$ & $2.962\times10^{-8}$ & $4.91\times10^{-3}$\\
2400 & $2.963\times10^{-7}$ & $2.964\times10^{-8}$ & $4.91\times10^{-3}$\\
4800 & $2.962\times10^{-7}$ & $2.964\times10^{-8}$ & $4.91\times10^{-3}$\\
\end{tabular}
\end{ruledtabular}
\end{table}

\begin{table}[!h]
\caption{\label{tab:app_true4l}Direct four-level optimization sanity check for
one representative seed.  This check supports the qualitative regime but is not
used for the main statistical claims.}
\begin{ruledtabular}
\begin{tabular}{lccc}
method & $1-F$ & 4L $\Pfinal$ & 4L $\Pbardeph$\\
\colrule
DRAG & $4.21{\times}10^{-5}$ & $4.18{\times}10^{-5}$ & $6.24{\times}10^{-3}$\\
DRAG$+$single-tone ECP & $7.30{\times}10^{-6}$ & $3.65{\times}10^{-8}$ & $5.98{\times}10^{-3}$\\
PSP & $1.96{\times}10^{-8}$ & $1.96{\times}10^{-8}$ & $5.13{\times}10^{-3}$\\
PSP$+$single-tone ECP & $1.94{\times}10^{-8}$ & $1.85{\times}10^{-8}$ & $5.13{\times}10^{-3}$\\
\end{tabular}
\end{ruledtabular}
\end{table}

The heuristic coupling
$w_M=0.1w_{\mathrm{path}}$ [Eq.~(\ref{eq:cmain})] places the operating
point on a broad plateau.  Sweeping the ratio
$w_M/w_{\mathrm{path}}\in\{0.01,\dots,1.0\}$ at $w_{\mathrm{path}}=30$
(three seeds, $n_{\mathrm{steps}}=1200$) changes the path-basin path reduction by
less than $1\%$ ($\Pbar=4.73$--$4.75\times10^{-3}$, a $0.62\%$ spread) and leaves
the path-shaping coefficient stable ($a_2=0.28$--$0.30$); the endpoint floor
stays at the numerical floor throughout, drifting only from ${\sim}10^{-10}$ to
${\sim}10^{-8}$ as the peak penalty is increased a hundredfold.  The prefactor is
therefore not a tuned parameter, and the sweep stays effectively single-parameter
in $w_{\mathrm{path}}$.
\FloatBarrier

% ============================================================================
\section{Regime and stress boundary}\label{app:stress}
% ============================================================================

Across the grid $T\in\{6,8,10,12,15\}$~ns and
$\eta/2\pi\in\{0.15,0.175,0.2,0.225,0.25\}$~GHz, evaluated at the
$w_{\mathrm{path}}=30$ operating point, the path-basin solutions reduce the
three-level population path by ${\sim}11$--$20\%$ wherever the optimizer finds the
path basin (Table~\ref{tab:regime}), and this reduction magnitude is consistent
across the grid and across independent seed sets.  The path-basin occupancy quoted in parentheses, by contrast, is a
property of the stochastic optimizer rather than of the method: with only a handful
of CMA-ES restarts it fluctuates from cell to cell, so in practice we run several
seeds and keep the best converged pulse.  Because $w_{\mathrm{path}}=30$ is selected at
the operating-point knee (Sec.~\ref{sec:sweep}) and that knee moves with the gate's
time--bandwidth budget, and because the ECP tones are matched to the anharmonicity,
a change of gate time or anharmonicity calls for re-selecting
$w_{\mathrm{path}}$ and recalibrating the endpoint tones rather than reusing
the operating-point values.

A second, distinct limit appears at the shortest-gate, smallest-anharmonicity
corner $(6~\mathrm{ns},0.15~\mathrm{GHz})$. There the bare PSP still
reduces the path by ${\sim}17\%$, but the coherent error is elevated
($1-F\approx5\times10^{-4}$) and, critically, completing the design forces the
single-tone ECP to spend a large out-of-band drive fraction
($C_{\mathrm{hf}}\approx0.11$), so the two-tone pulse no longer satisfies
the adopted waveform constraints. Increasing the anharmonicity restores
feasibility: at
$(6~\mathrm{ns},\,\eta/2\pi\ge0.2~\mathrm{GHz})$ the coherent error drops to the
${\sim}10^{-10}$ floor and $C_{\mathrm{hf}}$ returns to its ${\sim}10^{-5}$
baseline. Sufficient anharmonicity is thus a prerequisite for ultrafast
gates---a time--bandwidth--anharmonicity constraint.

\setcounter{table}{9}
\begin{table*}[t]
\caption{\label{tab:regime}Three-level path reduction relative to cosine DRAG
(PSP, path-basin median) across the regime grid, evaluated at the
$w_{\mathrm{path}}=30$ operating point. Each entry is the basin-median
reduction with the path-basin occupancy (out of five seeds) in parentheses; the
basin is taken as seeds reaching at least a $5\%$ reduction. The reduction is ${\sim}11$--$20\%$ wherever the basin is found.}
\begin{ruledtabular}
\begin{tabular}{lccccc}
$T$ (ns) $\backslash$ $\eta/2\pi$ (GHz) & 0.15 & 0.175 & 0.20 & 0.225 & 0.25\\
\colrule
6  & $17.4\%$ (5/5) & $16.2\%$ (5/5) & $18.4\%$ (5/5) & $14.7\%$ (3/5)  & $15.4\%$ (4/5)\\
8  & $18.2\%$ (2/5) & $16.1\%$ (3/5) & $15.0\%$ (4/5) & $11.5\%$ (2/5) & $18.5\%$ (3/5)\\
10 & $15.7\%$ (5/5) & $17.3\%$ (3/5) & $18.1\%$ (5/5) & $19.4\%$ (5/5) & $19.6\%$ (3/5)\\
12 & $13.3\%$ (3/5) & $17.8\%$ (4/5) & $18.5\%$ (2/5) & $18.2\%$ (5/5) & $17.1\%$ (3/5)\\
15 & $19.4\%$ (4/5) & $18.8\%$ (4/5) & $17.0\%$ (4/5) & $15.7\%$ (5/5) & $12.4\%$ (5/5)\\
\end{tabular}
\end{ruledtabular}
\end{table*}

% ============================================================================
\section{Robustness and hardware feasibility}\label{app:hardware}
% ============================================================================

\subsection{Peak drive amplitude}
\suppressfloats[t]

Path shaping does not increase the peak drive
strength.  In the three-level proxy at the operating point, the peak Rabi
rate of the PSP is $0.259$~rad\,ns$^{-1}$ versus $0.313$~rad\,ns$^{-1}$ for the
calibrated cosine-DRAG reference, an ${\approx}17\%$ reduction that is
identical across the five seeds, and adding the ECP tones changes the peak by
less than $0.1\%$.  The composite pulse therefore does not increase the peak-amplitude
requirement on the waveform generator; the out-of-band
fraction $C_{\mathrm{hf}}$ is an order of magnitude larger than for cosine DRAG
but remains at the $10^{-6}$ level.

\begin{figure}[!b]
\centering
\includegraphics[width=\columnwidth]{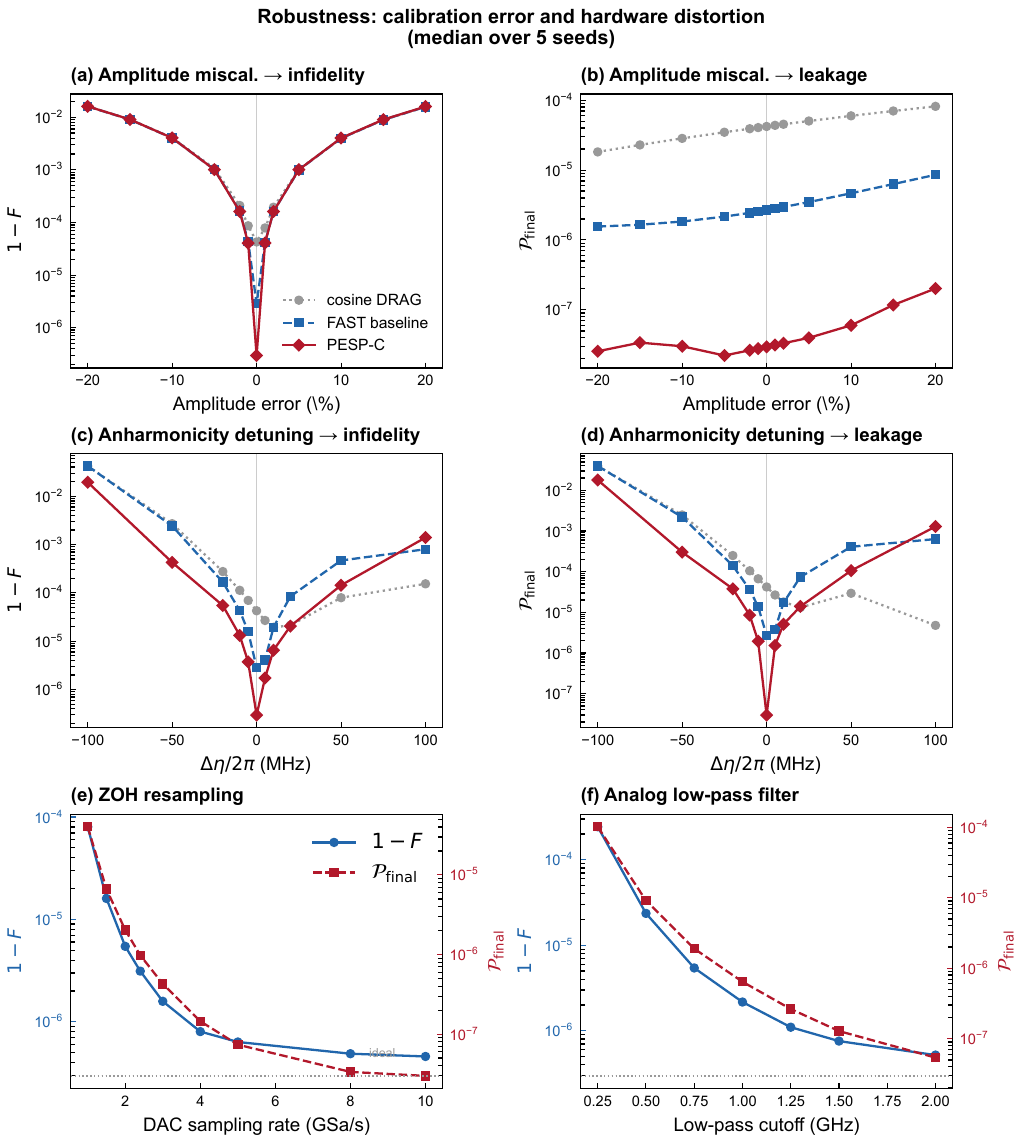}
\caption{\label{fig:robustness}Sensitivity of the PESP-C design to calibration and hardware imperfections (all-seed median over five seeds).
(a)~Amplitude miscalibration: the infidelity follows the coherent over-rotation
$\delta\theta^2/6$ while leakage remains at the $10^{-8}$ floor.
(b)~Anharmonicity detuning $\Delta\eta/2\pi$: the dominant sensitive axis for
endpoint cancellation, asymmetric (smaller $\eta$ is worse).
(c)~ZOH resampling: infidelity falls steeply with sampling rate; $\ge2.4$~GS/s
is adequate.
(d)~Analog low-pass filtering: recovery is possible via Wiener predistortion
for cutoffs $\ge0.5$~GHz.}
\end{figure}

\subsection{Amplitude and anharmonicity mismatch}
Table~\ref{tab:robust} (top) and Fig.~\ref{fig:robustness} report sensitivity
to amplitude and $\eta$ errors,
evaluated for the PESP-C design at the operating point
($w_{\mathrm{path}}=30$; ideal $1-F=3.0\times10^{-7}$,
$\Pfinal=3.0\times10^{-8}$, all-seed median). The amplitude-error infidelity is
dominated by the coherent over-rotation $\delta\theta^2/6$
($\pm1\%\to4.0$--$4.1\times10^{-5}$ for the $\pi/2$ rotation), which is
operating-point independent and removable by a Rabi amplitude / virtual-$Z$
recalibration; the calibration-independent leakage remains at the
${\sim}3\times10^{-8}$ floor at $\pm1\%$---the second tone maintains endpoint
cancellation under moderate amplitude mismatch. The $\eta$ mismatch is the
dominant sensitive axis for endpoint cancellation, since the auxiliary-tone
match depends on the anharmonicity, and is asymmetric (a smaller anharmonicity
is worse); a $\pm20$~MHz mismatch costs ${\sim}2$--$5\times10^{-5}$. This
implies a periodic recalibration cadence, echoing the ALC calibration sequence
of Ref.~\cite{Chiaro2025}.

\subsection{Sampling and filtering}
Table~\ref{tab:robust} (bottom) reports the effect of zero-order-hold (ZOH)
sampling and analog low-pass (LP) filtering for the PESP-C design at the operating point (ideal $1-F=3.0\times10^{-7}$). Linear low-pass distortion is a
linear filter and is recoverable close to the ideal floor by Wiener
predistortion ($\epsilon=10^{-3}$); raw, it costs $2.4\times10^{-4}$ at
$0.25$~GHz, falling to ${\sim}2\times10^{-6}$ by $1$~GHz. ZOH sampling at
$1$~GS/s ($8.0\times10^{-5}$) is not linearly recoverable, but the penalty falls
steeply with rate---${\sim}3\times10^{-6}$ at $2.4$~GS/s and ${\sim}6\times10^{-7}$
(near ideal) at $5$~GS/s---so there is a clear sampling-rate threshold.

\FloatBarrier

\setcounter{table}{10}
\begin{table}[!h]
\caption{\label{tab:robust}Hardware sensitivity of the PESP-C design at the operating point ($w_{\mathrm{path}}=30$, all-seed
median).  Top: amplitude and anharmonicity calibration errors (the amplitude
$1-F$ is the calibration-removable over-rotation $\delta\theta^2/6$; the leakage
column is calibration-independent).  Bottom: zero-order-hold (ZOH) sampling and
analog low-pass (LP) filtering, before predistortion.}
\begin{ruledtabular}
\begin{tabular}{lcc}
setting / error & $1-F$ & $\Pfinal$\\
\colrule
ideal (oper.\ pt.) & $3.0\times10^{-7}$ & $3.0\times10^{-8}$\\
amp $-1\%$        & $4.0\times10^{-5}$ & $2.8\times10^{-8}$\\
amp $+1\%$        & $4.1\times10^{-5}$ & $3.1\times10^{-8}$\\
$\eta-5$~MHz      & $3.7\times10^{-6}$ & $1.9\times10^{-6}$\\
$\eta+5$~MHz      & $1.7\times10^{-6}$ & $1.5\times10^{-6}$\\
$\eta-20$~MHz     & $5.5\times10^{-5}$ & $3.8\times10^{-5}$\\
$\eta+20$~MHz     & $2.0\times10^{-5}$ & $1.4\times10^{-5}$\\
\colrule
ZOH 1 GS/s     & $8.0\times10^{-5}$ & $4.0\times10^{-5}$\\
ZOH 2.4 GS/s   & $3.1\times10^{-6}$ & $9.8\times10^{-7}$\\
ZOH 5 GS/s     & $6.3\times10^{-7}$ & $7.4\times10^{-8}$\\
LP 0.25 GHz     & $2.4\times10^{-4}$ & $1.0\times10^{-4}$\\
LP 0.5 GHz      & $2.4\times10^{-5}$ & $9.1\times10^{-6}$\\
LP 1.0 GHz      & $2.2\times10^{-6}$ & $6.4\times10^{-7}$\\
\end{tabular}
\end{ruledtabular}
\end{table}

\FloatBarrier


\begin{thebibliography}{99}

\bibitem{Koch2007}
J.~Koch, T.~M. Yu, J.~Gambetta, A.~A. Houck, D.~I. Schuster, J.~Majer,
A.~Blais, M.~H. Devoret, S.~M. Girvin, and R.~J. Schoelkopf,
Charge-insensitive qubit design derived from the Cooper pair box,
Phys. Rev. A \textbf{76}, 042319 (2007).

\bibitem{Blais2021}
A.~Blais, A.~L. Grimsmo, S.~M. Girvin, and A.~Wallraff,
Circuit quantum electrodynamics,
Rev. Mod. Phys. \textbf{93}, 025005 (2021).

\bibitem{Krantz2019}
P.~Krantz, M.~Kjaergaard, F.~Yan, T.~P. Orlando, S.~Gustavsson, and
W.~D. Oliver,
A quantum engineer's guide to superconducting qubits,
Appl. Phys. Rev. \textbf{6}, 021318 (2019).

\bibitem{Kjaergaard2020}
M.~Kjaergaard, M.~E. Schwartz, J.~Braum\"uller, P.~Krantz, J.~I.-J. Wang,
S.~Gustavsson, and W.~D. Oliver,
Superconducting qubits: Current state of play,
Annu. Rev. Condens. Matter Phys. \textbf{11}, 369 (2020).

\bibitem{Motzoi2009}
F.~Motzoi, J.~M. Gambetta, P.~Rebentrost, and F.~K. Wilhelm,
Simple pulses for elimination of leakage in weakly nonlinear qubits,
Phys. Rev. Lett. \textbf{103}, 110501 (2009).

\bibitem{Chen2016}
Z.~Chen \textit{et al.},
Measuring and suppressing quantum state leakage in a superconducting
qubit, Phys. Rev. Lett. \textbf{116}, 020501 (2016).

\bibitem{Wood2018}
C.~J. Wood and J.~M. Gambetta,
Quantification and characterization of leakage errors,
Phys. Rev. A \textbf{97}, 032306 (2018).

\bibitem{Fowler2012}
A.~G. Fowler, M.~Mariantoni, J.~M. Martinis, and A.~N. Cleland,
Surface codes: Towards practical large-scale quantum computation,
Phys. Rev. A \textbf{86}, 032324 (2012).

\bibitem{McEwen2021}
M.~McEwen \textit{et al.},
Removing leakage-induced correlated errors in superconducting quantum error correction,
Nat. Commun. \textbf{12}, 1761 (2021).

\bibitem{Miao2023}
K.~C. Miao \textit{et al.},
Overcoming leakage in quantum error correction,
Nat. Phys. \textbf{19}, 1780 (2023).

\bibitem{Strauch2025}
F.~W. Strauch,
Dephasing-induced leakage in multilevel superconducting quantum
circuits, Phys. Rev. A \textbf{112}, 012601 (2025).

\bibitem{Gambetta2011}
J.~M. Gambetta, F.~Motzoi, S.~T. Merkel, and F.~K. Wilhelm,
Analytic control methods for high-fidelity unitary operations in a
weakly nonlinear oscillator, Phys. Rev. A \textbf{83}, 012308 (2011).

\bibitem{Lucero2010}
E.~Lucero, J.~Kelly, R.~C. Bialczak, M.~Lenander, M.~Mariantoni, M.~Neeley,
A.~D. O'Connell, D.~Sank, H.~Wang, M.~Weides, J.~Wenner, T.~Yamamoto,
A.~N. Cleland, and J.~M. Martinis,
Reduced phase error through optimized control of a superconducting
qubit, Phys. Rev. A \textbf{82}, 042339 (2010).

\bibitem{Hyyppa2024}
E.~Hyypp\"a \textit{et al.},
Reducing leakage of single-qubit gates for superconducting quantum
processors using analytical control pulse envelopes,
PRX Quantum \textbf{5}, 030353 (2024).

\bibitem{Chiaro2025}
B.~Chiaro and Y.~Zhang,
Active leakage cancellation in single qubit gates,
Phys. Rev. Lett. \textbf{135}, 130601 (2025).

\bibitem{Khaneja2005}
N.~Khaneja, T.~Reiss, C.~Kehlet, T.~Schulte-Herbr\"uggen, and S.~J. Glaser,
Optimal control of coupled spin dynamics: design of NMR pulse
sequences by gradient ascent algorithms,
J. Magn. Reson. \textbf{172}, 296 (2005).

\bibitem{Werninghaus2021}
M.~Werninghaus, D.~J. Egger, F.~Roy, S.~Machnes, F.~K. Wilhelm, and S.~Filipp,
Leakage reduction in fast superconducting qubit gates via optimal
control, npj Quantum Inf. \textbf{7}, 14 (2021).

\bibitem{Caneva2011}
T.~Caneva, T.~Calarco, and S.~Montangero,
Chopped random-basis quantum optimization,
Phys. Rev. A \textbf{84}, 022326 (2011).

\bibitem{Glaser2025}
N.~Glaser \textit{et al.},
Closed-loop optimization for high-fidelity controlled-$Z$ gates in
superconducting qubits, Phys. Rev. Applied \textbf{24}, 024048 (2025).

\bibitem{Nguyen2024}
H.~N. Nguyen, F.~Motzoi, M.~Metcalf, K.~B. Whaley, M.~Bukov, and M.~Schmitt,
Reinforcement learning pulses for transmon qubit entangling gates,
Mach. Learn.: Sci. Technol. \textbf{5}, 025066 (2024).

\bibitem{McKay2017}
D.~C. McKay, C.~J. Wood, S.~Sheldon, J.~M. Chow, and J.~M. Gambetta,
Efficient $Z$ gates for quantum computing,
Phys. Rev. A \textbf{96}, 022330 (2017).

\bibitem{Wang2025}
R.~Wang \textit{et al.},
Suppressing spurious transitions using spectrally balanced pulse,
Phys. Rev. Lett. \textbf{135}, 160804 (2025).

\bibitem{Li2024}
B.~Li, T.~Calarco, and F.~Motzoi,
Experimental error suppression in cross-resonance gates via multi-derivative pulse shaping,
npj Quantum Inf. \textbf{10}, 66 (2024).

\bibitem{Cywinski2008}
\L.~Cywi\'nski, R.~M. Lutchyn, C.~P. Nave, and S.~Das Sarma,
How to enhance dephasing time in superconducting qubits,
Phys. Rev. B \textbf{77}, 174509 (2008).

\bibitem{Green2013}
T.~J. Green, J.~Sastrawan, H.~Uys, and M.~J. Biercuk,
Arbitrary quantum control of qubits in the presence of universal
noise, New J. Phys. \textbf{15}, 095004 (2013).

\bibitem{PazSilva2014}
G.~A. Paz-Silva and L.~Viola,
General transfer-function approach to noise filtering in open-loop
quantum control, Phys. Rev. Lett. \textbf{113}, 250501 (2014).

\bibitem{Oda2023}
Y.~Oda, D.~Lucarelli, K.~Schultz, B.~D. Clader, and G.~Quiroz,
Optimally band-limited noise filtering for single-qubit gates,
Phys. Rev. Applied \textbf{19}, 014062 (2023).

\bibitem{Poggi2025}
P.~M. Poggi and A.~Kiely,
Suppressing leakage and maintaining robustness in transmon qubits:
Signatures of a trade-off relation,
Control Eng. Pract. \textbf{172}, 106855 (2026).

\bibitem{McCord2025}
J.~J. McCord, M.~Kuzmanovi\'c, and G.~S. Paraoanu,
Pareto-optimality of pulses for robust population transfer in a ladder-type qutrit,
EPJ Quantum Technol. \textbf{12}, 121 (2025).

\end{thebibliography}
\end{document}